\documentclass[11pt, twoside]{article}
\usepackage{amsmath, amssymb, amsthm}
\usepackage{bm, bbm}
\usepackage{algorithm}
\usepackage{algpseudocode}
\usepackage{float, graphicx, subcaption, setspace, multicol}
\usepackage{comment}
\usepackage{url}
\usepackage{enumitem}
\usepackage{hyperref}
\usepackage{natbib}
\usepackage[usenames,dvipsnames]{xcolor}
\usepackage{nicematrix}
\usepackage{csquotes}
\usepackage{caption}
\usepackage{cleveref}
\usepackage{multicol}

\definecolor{SkyBlue}{RGB}{14, 118, 188}
\definecolor{BrightRed}{RGB}{223, 82, 78}
\definecolor{Green638}{RGB}{165,255,118} % from colours.cafe on instagram; pallete638

\hypersetup{pdfborder = {0 0 0.5 [3 3]}, colorlinks = true, linkcolor = BrightRed, citecolor = SkyBlue}

\newcommand{\E}{\mathbb{E}} % boldfaced E for expectations
\def\P{\mathbb{P}} % boldfaced P for probability. overriding \P for paragraph symbol

\theoremstyle{plain}

\theoremstyle{definition}

\newtheorem{ex}{Example}

\theoremstyle{remark}

\newcommand{\contextshort}{\mathcal{G}}

\newcommand{\personelshort}{\mathcal{P}}

\newcommand{\locationshort}{\mathcal{L}}

\newcommand{\swing}{\texttt{swing}}
\newcommand{\contact}{\texttt{contact}}
\newcommand{\strike}{\texttt{strike}}
\newcommand{\outcome}{\texttt{outcome}}
\newcommand{\runs}{R}
\newcommand{\correct}{\texttt{xR\_optimal}}

\newcommand{\gamestatecat}{\texttt{gstate}}

\newcommand{\bartxr}{\texttt{BARTxR}}
\newcommand{\rex}{\texttt{REx}}
\newcommand{\bayesrex}{\texttt{BayesREx}}
\newcommand{\fgre}{\texttt{RE24}}
\newcommand{\evdiff}{\texttt{EVdiff}}

\usepackage{fullpage, parskip} 

\title{Evaluating plate discipline in Major League Baseball with Bayesian Additive Regression Trees}
\author{Ryan Yee\thanks{Department of Statistics, University of Wisconsin--Madison. \url{ryee2@wisc.edu}} \and Sameer K.~Deshpande\thanks{Department\ of Statistics, University of Wisconsin--Madison. \url{sameer.deshpande@wisc.edu}}}
\def\codelinkblind{\texttt{link to GitHub repository blinded for review}}
\def\codelinkunblind{\url{https://github.com/ryanyee3/plate_discipline_code}}

\def\shinylinkblind{\texttt{link to GitHub repository blinded for review}}
\def\shinylinkunblind{\url{https://ryanyee3.shinyapps.io/batter_evaluation_app/}}

\def\blind{0}

\def\codelink{
  \if\blind1
  \codelinkblind
  \else 
  \codelinkunblind
  \fi
}

\def\shinylink{
  \if\blind1
  \shinylinkblind
  \else 
  \shinylinkunblind
  \fi
}

\begin{document}
\maketitle

We introduce a three-step framework to determine at which pitches Major League batters should swing. 
Unlike traditional plate discipline metrics, which implicitly assume that all batters should always swing at (resp. take) pitches inside (resp. outside) the strike zone, our approach explicitly accounts not only for the players and umpires involved in the pitch but also in-game contextual information like the number of outs, the count, baserunners, and score. 
We first fit flexible Bayesian nonparametric models to estimate (i) the probability that the pitch is called a strike if the batter takes the pitch; (ii) the probability that the batter makes contact if he swings; and (iii) the number of runs the batting team is expected to score following each pitch outcome (e.g. swing and miss, take a called strike, etc.). 
We then combine these intermediate estimates to determine whether swinging increases the batting team's run expectancy. 
Our approach enables natural uncertainty propagation so that we can not only determine the optimal swing/take decision but also quantify our confidence in that decision. 
We illustrate our framework using a case study of pitches faced by Mike Trout in 2019.

\onehalfspacing
\section{Introduction}
\label{sec:intro}
\subsection{Motivating example}
\label{subsection:motivation}

\begin{figure}[ht]
    \centering
    \includegraphics[width=0.9\textwidth]{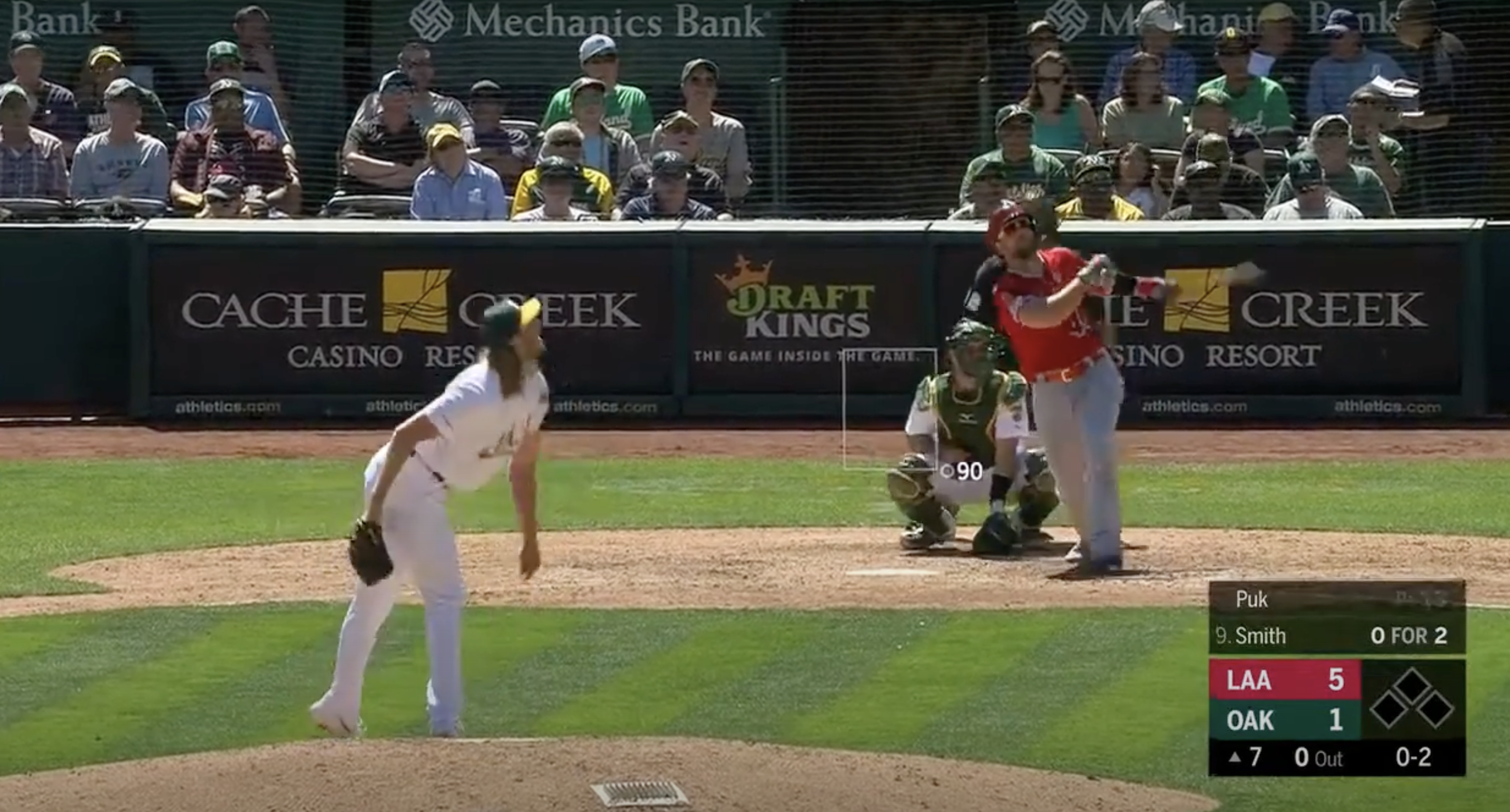}
    \caption[short]{Kevan Smith swings at a fastball thrown just outside the lower right-hand corner of the strike zone.}
    \label{fig:example}
\end{figure}

At the top of the seventh inning of the September 5, 2019 game between the Oakland Athletics and the Los Angeles Angels, Angels batter Kevan Smith faced Athletics pitcher A.J.\ Puk with the Angels leading 5 runs to 1.
On an 0-2 pitch with no outs and no runners on-base, Puk threw a fastball that just missed the lower right-hand corner of the strike zone (see \Cref{fig:example}). 
Smith decided to swing at the pitch.
%Had Smith not swung at the pitch and had the pitch been called a ball, it would have extended the at-bat, leading to a more favorable 1-2 count.
%Instead, Smith decided to swing at the pitch.
Did Smith make the correct decision?

By swinging, Smith risked missing the pitch, picking up a strike, and losing an out, thereby putting his team at a disadvantage.
And even if he had made contact, he risked flying or grounding out, which would also disadvantage his team.
At the same time, however, by swinging, Smith had a chance of getting on base or even scoring a home run.
On the other hand, had Smith not swung and had instead taken the pitch, the umpire may have correctly called the pitch a ball (which would advantage Smith's team) or mistakenly called a strike (which would disadvantage Smith's team).

% image from 1:03 --> https://www.youtube.com/watch?v=mTFfAJW7WJ8

As it turns out, Smith hit a home run on this pitch.
So at least retrospectively, it would seem like the decision to swing was good.
%However, determining whether Smith made the right decision \textit{prior to observing the outcome} is a substantially harder problem 
But was Smith simply lucky?
Or should we expect him to hit home runs consistently off of similar pitches in similar game situations? 
And how sure should we be about these expectations?
%That is, should we expect him to hit home runs consistently off of similar pitches in similar game situations?

We provide quantitative answers to these questions using a Bayesian modeling framework to assess batter's decision making --- or plate discipline --- \textit{before} observing the outcome of each pitch. 
In the case of Smith, we find that there was a 12.3\% chance (90\% credible interval [5.8\%, 16.7\%]) that the umpire would call the pitch a strike.
Although taking the pitch would likely have benefitted Smith's team, Smith had about an equal chance of making contact if he swung (83.1\%, 90\% credible interval $[80.6\%, 85.3\%]$).
In fact, we find that by swinging on similar pitches, Smith increases the number of runs the Angels are expected to score in the remainder of the half-inning only 61.4\% of the time and only by about 0.01 runs on average (90\% credible interval $[-0.04, 0.08]$).
Ultimately, our analysis concludes that while Smith's decision to swing on the pitch was not especially risky, it was unlikely to substantially benefit his team, in terms of the number of expected runs the team would score in the rest of the inning. 

Traditional plate discipline metrics compare the proportion of pitches batters swing at that are outside and inside the strike zone \citep{fangraphs_PD_metrics}.
In this way, these metrics reward batters for avoiding taken strikes based on the implicit assumption that pitches thrown outside (resp.\ inside) the strike zone are always called balls (resp.\ strikes).
Of course, umpires do not adhere strictly to the official strike zone when making ball/strike decisions.
Though they often rely on adaptive heuristics and prior experience to make calls \citep{GreenDaniels2021_instinct}, umpires can also be influenced by the framing ability of individual catchers \citep{Marchi-2011, lindbergh-2013, DeshpandeWyner2017}; player age or ability \citep{KimKing2014, Mills2014-hd}; their previous calls \citep{Chen2016}; and the fact that their calls are reviewed by the league \citep{Mills2017}.
As a result, traditional plate discipline metrics may systematically penalize batters who swing at ``frameable'' pitches that just barely miss the strike zone but are likely to be called strikes.
These metrics additionally fail to account for the fact that some pitches are easier to hit than others.
At least intuitively, we might regard a batter who only takes hard-to-hit pitches as more disciplined than a batter who consistently takes easy-to-hit pitches.
Further, traditional plate discipline metrics entirely ignore the many contextual and situational factors that can influence a batter's decision to swing.
For instance, batters may be more aggressive on pitches thrown with two strikes to avoid striking out while looking.
Finally, existing plate discipline metrics do not quantify the uncertainty in their findings.

\subsection{Our contributions}
\label{subsection:contribution}

We introduce a three-step approach to determine optimal swing decisions, assess batter decision-making on a per-pitch basis, and quantify our uncertainty about both.
In the first step, we fit flexible Bayesian nonparametric models that enable us to estimate, for any single pitch, (i) the probability that a batter makes contact; (ii) the probability that the umpire calls a strike; and (iii) the average number of runs that the batting team is expected to score in the remaining half-inning as functions of the pitch location, players and umpires involved, and game-state information like the count, inning, and baserunners. 
Then, in the second step, we combine these estimates using the law of total expectation to compute the expected number of runs the batting team is expected to score if the batter swings or takes the pitch.
Finally, we use these quantities to determine the optimal swing/take decision.
By fitting Bayesian models in the first step, we can propagate uncertainty about the actual outcomes of a particular pitch through to our assessment of the batter's decision making in an intuitive and computationally efficient manner.

Our framework involves fitting three component models, each of which may be of independent interest.
First, elaborating on work begun in \citet{DeshpandeWyner2017}, we develop a Bayesian model of called strike probability that ``borrows statistical strength'' across umpires and players through flexible partial pooling of data.
We demonstrate that this model, which is based on Bayesian additive regression trees \citep[BART;][]{Chipman2010} and accounts not only for pitch location but also player and umpire identities and game-state information, outperforms parametric and nonparametric competitors that only account for location.
We develop a similar model for the probability that a batter makes contact on a pitch.
Finally, we developed a BART-based run expectancy model, which we call $\bartxr,$ to predict the number of runs a team is expected to score following each pitch outcome as a function of in-game contextual variables like the count, number of outs, score differential, and baserunners.
At a high-level $\bartxr$ generalizes existing run expectancy measures like $\fgre$ \citep{tango2007book}, which computes the average number of runs scored within bins defined by the number of outs and configuration of baserunners.
Using a comprehensive cross-validation study, we demonstrate that $\bartxr$'s predictions are more accurate than those of $\fgre$ and several variants thereof.

The remainder of this paper is organized as follows.
We introduce the data and notation used in \Cref{subsection:data_notation} before describing our three-step framework for assessing batter decision-making in \Cref{subsection:batter_eval}.
Then, we detail our modeling approach in \Cref{sec:modeling}.
In \Cref{subsection:case_study}, we perform a case study of a single batter, Mike Trout, highlighting the types of plate discipline assessments that can be done using our framework.
We conclude in \Cref{sec:discussion} with a discussion of several extensions of our modeling framework.

\section{Data and background}
\label{sec:data_background}
\subsection{Data and notation}
\label{subsection:data_notation}

Our analysis uses pitch-by-pitch tracking data from Major League Baseball's Statcast database.
For each pitch, we observe game-state information (e.g.\ count, outs, score, baserunners), pitch personnel (e.g.\ pitcher, batter, fielders, umpire), pitch outcome (e.g.\ hit, ball, strike), and other game actions (e.g.\ steal, substitution).
Additionally, we observe the horizontal and vertical coordinates of each pitch's trajectory as it crosses the front edge of home plate. 
We scraped these data using the \textbf{baseballr} package \citep[version 1.2.0;][]{baseballr_package}.
We limited our analysis to pitches thrown during regular season games.
We fit our strike probability model using all 380,654 pitches that were taken during the 2019 season and we fit our contact probability model using all 341,725 pitches at which batters swung in the 2019 season.
We fit our expected runs model ($\bartxr$) using all 2,853,912 pitches thrown between 2015 and 2018, inclusive.
Observe that these three datasets are disjoint.

We use $\contextshort$ to denote information about the state of the game when the pitch was thrown including the count, outs, baserunners, score differential, inning, and whether it is the top or bottom of the inning.
We similarly use $\personelshort$ to record the personnel involved in a pitch including the identities of the batter, catcher, pitcher, and home plate umpire.
We additionally include indicators of the batter and pitcher handedness in $\personelshort$ as well as quantitive measures of the batter and pitcher quality, which we describe below.
Finally, we denote the location of the pitch (i.e.\ the \texttt{plate\_x} and \texttt{plate\_z} coordinates in Statcast) as its crosses the front edge of home plate with $\locationshort.$

Let $\swing=1$ if the batter swings and $\swing=0$ if the batter takes the pitch.  
If the batter decides to swing, let $\contact=1$ if the batter makes contact with the pitch and $\contact=0$ if the batter misses. 
If the batter does not swing, let $\strike=1$ if the umpire calls a strike and $\strike=0$ if the umpire calls a ball.
Let $\outcome=(\contact, \strike)$ be a vector denoting the outcome of the pitch where $\contact=\texttt{NA}$ if $\swing=0$ and $\strike=\texttt{NA}$ if $\swing=1$.
Let $\gamestatecat$ denote the game-state category the game moves to after the outcome of the pitch is observed.
$\gamestatecat$ encodes similar information as $\outcome$ but treats a called strike and a miss as the same game-state.

We quantify batter and pitcher quality using a running estimate of their weighted on-base average \cite[wOBA;][]{tango2007book}.
For each game, we set these quality measures to be their mean wOBA, averaged over all of their previous games in the season.
%For pitchers, this is the wOBA achieved from the results of all the pitches they have thrown up to the current game
For batters, higher wOBA represents higher quality while for pitchers, lower wOBA represents higher quality.
Note that for the first game, we set each player's quality measure to be the league average wOBA from the previous season.
Our choice to track player quality using a running wOBA  estimates follows the example of \citet{Brill2022-tp}.

\subsection{Determining the optimal decision}
\label{subsection:batter_eval}

To motivate our three-step framework, consider a pitch at the moment just before the batter decides to swing.
Suppose that we knew the number of runs the batter's team would score in the remainder of the half-inning if (a) the batter swings or (b) the batter does not swing.
Based on that knowledge, the optimal decision is intuitively the one that leads to scoring more runs.
Of course, we are uncertain about these run values at the moment just before the batter swings or takes the pitch.
This is because there is uncertainty in the ultimate outcome: if he swings, the batter might make contact or miss and if he takes the pitch, the umpire may call it a ball or a strike. 
Because of this uncertainty, we determine the optimal swinging decision using the \textit{expected} number of runs the batter's team will subsequently score.
That is, we will average over our uncertainty about these outcomes.

\begin{figure}[ht]
	\centering
	\includegraphics[width=6in]{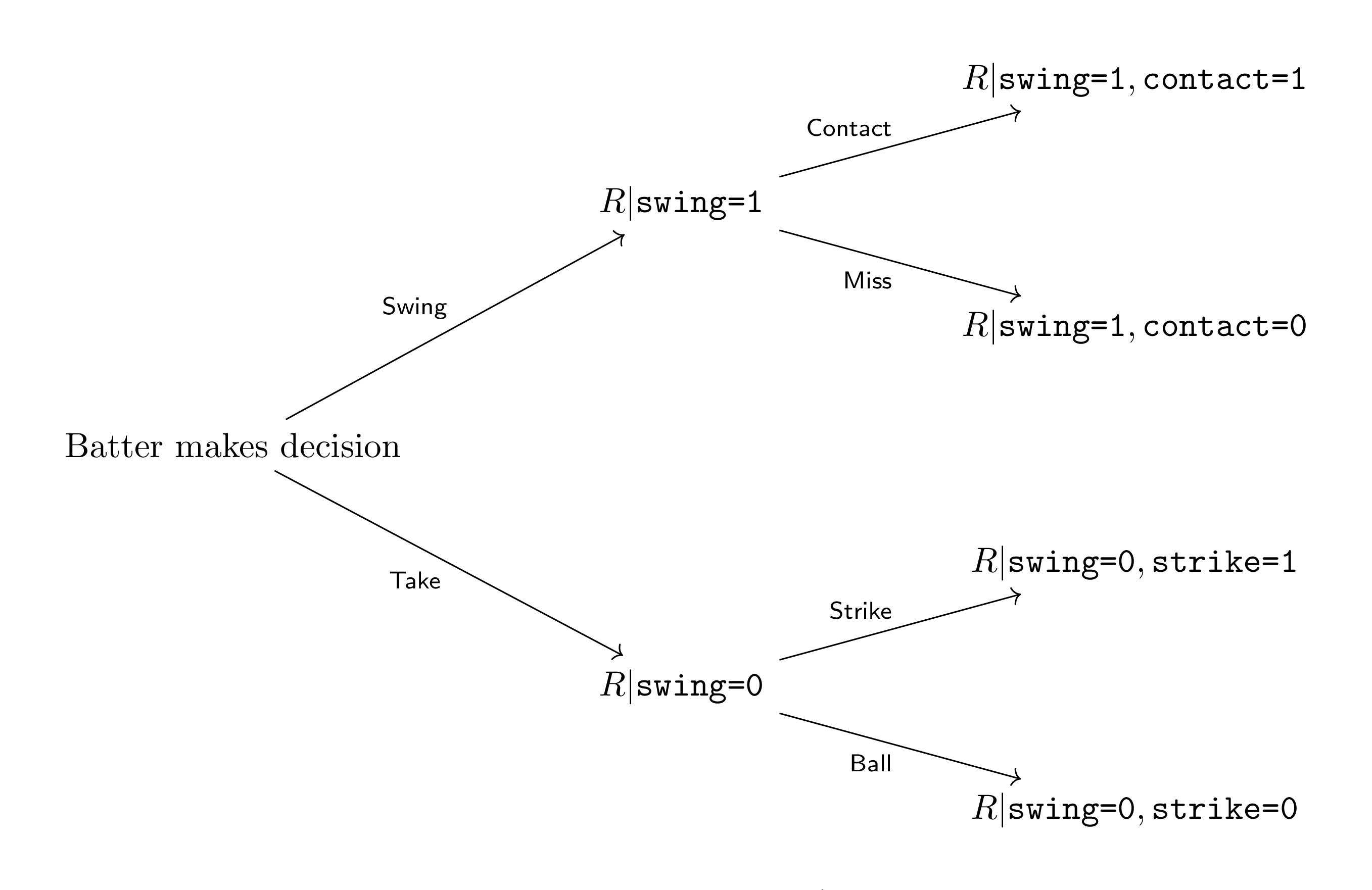}
	\caption{Framework for modeling the outcomes of a pitch.}
	\label{fig:framework}
\end{figure}

\Cref{fig:framework} illustrates the four possible outcomes following the batter's decision.
Computing the expected number of runs scored after a swing requires knowledge of (i) the probability the batter will make contact; (ii) the expected number of runs the batting team will score in the current half-inning after making contact; and (iii) the expected number of runs he batting team will score in the current half-inning after swinging but missing.
Similarly, computing the expected number of runs scored after a take requires knowledge of (i) the probability the umpire will call a strike; (ii) the expected number of runs the batting team will score in the current half-inning after a called strike; and (iii) the expected number of runs the batting team will score in the current half-inning after a called ball.

Formally, let $\runs$ be the number of runs the batting team scores following the pitch.
Given the game-state information $\contextshort$, personnel $\personelshort$, and location $\locationshort$ of a pitch, we need to compute $\E[\runs \vert \contextshort, \personelshort, \locationshort, \swing]$ for both values of $\swing$ in order to determine the optimal decision for that pitch.
Observe that we can decompose the expected runs following a swing or take as
\begin{equation} 
\label{eq:xR} 
\mathbb{E}[\runs \vert \contextshort, \personelshort, \locationshort, \swing] = \sum\limits_{\outcome} \mathbb{P}(\outcome \vert \contextshort, \personelshort, \locationshort, \swing) \mathbb{E}[\runs \vert \contextshort, \personelshort, \locationshort, \swing, \outcome].
\end{equation} 

To determine the optimal decision, we introduce $\evdiff$ as the difference in expected runs following a swing and following a take.
Formally, we define
\begin{equation} 
\label{eq:evdiff} 
\evdiff = \mathbb{E}[\runs \vert \contextshort, \personelshort, \locationshort, \swing = 1] - \mathbb{E}[\runs \vert \contextshort, \personelshort, \locationshort, \swing = 0]. 
\end{equation}
The sign of $\evdiff$ determines the optimal decision $\correct$, which is given by
$$
\correct = 
		\begin{cases}
			\swing = 1, & \text{if $\evdiff > 0$} \\
			\swing = 0, & \text{if $\evdiff \leq 0.$}
		\end{cases}
$$

Determining the optimal decision for a pitch and subsequently assessing a batter's decision-making requires knowledge of $\mathbb{E}[\runs \vert \contextshort, \personelshort, \locationshort, \swing, \outcome]$ and $\P(\outcome \vert \contextshort, \personelshort, \locationshort, \swing)$. 
Although we do not know these quantities exactly, we can estimate them using our collected data. 
We will discuss each estimation problem in Section~\ref{sec:modeling}.

\subsection{Related work and background on BART}
\label{sec:related_work_bart}

\textbf{Existing plate discipline metrics}.
Traditional plate discipline metrics \citep[e.g.,][]{fangraphs_PD_metrics} characterize how often batters swing at pitches thrown outside (\texttt{O-Swing\%}) and inside (\texttt{Z-Swing\%}) the strike zone as well as how often batters made contact with these pitches (\texttt{O-Contact\%} and \texttt{Z-Contact\%}, respectively).
Intuitively, disciplined batters have high \texttt{Z-Swing\%} and low \texttt{O-Swing\%}.
Unfortunately, such metrics implicitly assume that all pitches thrown inside (resp.\ outside) the strike zone are equally desirable to hit (resp.\ not hit) which can give a false impression of a batter's plate discipline.
For example, Kevan Smith's decision in \Cref{fig:example} would negatively affect his \texttt{O-Swing\%} even though he had a high probability of making contact.
By treating every pitch equally, traditional metrics fail to account for important contextual factors that impact a batter's decision to swing such as strike probability \citep[see, e.g.,][]{moonshot-sz, Green2014-pj, Mills2014-hd}.
As a result, they may systematically over-penalize batters who see many pitches near the edges of the strike zone.

To overcome these limitations, recent studies have focused on directly modeling the batter's actual decision-making process.
For instance, \citet{moonshot-pd} first estimates the probability that each pitch is called a strike and then uses those estimates to predict whether a batter will swing at a pitch.
The authors characterize disciplined batters as those whose swing probabilities increase most in response to small increases in strike probability. 
Unfortunately, this characterization does not consider the downstream impact swing decisions have on metrics like run expectancy or win probability.
Towards this end, \citet{VockVock2018} used a causal inference framework to predict how a batter's batting average, on-base percentage, and slugging percentage would change under different counterfactual decision-making strategies.
At a high-level, their framework answers the question ``What would happen if batter A made swing/take decisions like batter B?''
While interesting and informative, the models in \citet{moonshot-pd} and \citet{VockVock2018} make no attempt to determine the decision a batter \textit{ought} to make.
In the context of \Cref{fig:framework}, these models try to predict which path a batter will follow.
Our approach, in contrast, attempts to determine which path would most benefit the batter's team. 

Independently of but concurrently to this work, \citet{eagel-p1,eagel-p2} introduced the Expected Additional runs Gained by Looking/swinging Estimate (EAGLE) model to quantify plate discipline.
Like us, they also used a tree-structured framework to determine the optimal swing/take decision and fit intermediate strike probability and contact probability models using flexible, non-parametric procedures.
Superficially, our approach differs from theirs in terms of the predictor variables included in these intermediate probability models as well as the specific model fitting procedure used (BART in our case versus gradient-boosted trees in theirs). 
Like us, EAGLE combines predictions from intermediate probability models with a run expectancy model to compute the expected number of runs a team will score following a swing or take.
EAGLE uses a variant of \texttt{RE24} that accounts for baserunners, count, and outs\footnote{We include this model in our cross-validation study in \Cref{subsection:model_validation} as \texttt{RE288}} followed by an \textit{ad hoc} correction for batter quality.
Our approach, on the other hand, uses predictions from a much higher-resolution regression-based run expectancy model.

The most substantive difference between our proposal and EAGLE's lies in uncertainty quantification.
Simply put, EAGLE makes no attempt to propagate uncertainties about the intermediately-estimated strike probability, contact probability, and run expectancies to their evaluation of batter decision making.
In sharp contrast, our Bayesian approach makes such uncertainty propagation easy (see \Cref{subsection:uncertainty_propagation}). 
We argue that such uncertainty quantification is of paramount importance for evaluating decision-making.
Basically, we do not wish to penalize batters for making suboptimal decisions when there is considerable uncertainty about what the optimal decision even is!

% EAGLE Approach
% (i) compute the difference in pre and post game run expectancy (DRE) for each pitch outcome using RE288, adjust based on who the batter is
% (ii) calculate strike probability regressing plate_x, plate_z, top, bottom of strike zone using XGBoost
% (iii) calculate swing outcome probabilities regressing velocity, spin rate, movement, location, pitch type, count, outs, handedness of pitcher/batter, batter characteristics, pitcher characteristics -- confusion matrix for this is not great
% (iv) calculate xR from swinging and taking in the same way we do

%Our approach is most similar to the concurrently developed EAGLE (Expected Additional runs Gained by Looking/swinging Estimate) metric \citep{eagel-p1,eagel-p2}, which 

\textbf{BART}. Initially introduced in the context of nonparametric regression, BART has emerged as a popular ``off-the-shelf'' modeling tool because it often delivers accurate predictions with reasonably well-calibrated uncertainty estimates without requiring users to (a) pre-specify the parametric form of the regression function and (b) tune any hyperparameters. 
At a high-level, BART works by approximating unknown functions with sums of binary regression trees and excels at capturing complicated nonlinearities and complex, high-order interaction effects. 
We believe \textit{a priori} that both strike and contact probabilities are highly non-linear and may depend on complicated interactions between players, umpires, pitch location, and in-game contextual variables like the count or baserunner configuration. 
Insofar as such nonlinearities and interactions are difficult to specify correctly in a parametric fashion, BART is an especially attractive modeling choice as it does not require us to pre-specify such a parametric form.
We fit our three BART models using the \textbf{flexBART} package, which permits more flexible modeling with categorical predictors such as batter identity that can take on many values.
See \citet{Deshpande2022} for more details.

\section{Modeling and uncertainty propagation}
\label{sec:modeling}
\subsection{Uncertainty propagation}
\label{subsection:uncertainty_propagation}

Recall that computing $\evdiff$ requires first estimating each term in the summand in \Cref{eq:xR}.
By taking a Bayesian approach, we can quantify how uncertainties about these estimates propagate to uncertainty about $\evdiff$ and $\correct$ in a relatively straightforward fashion.
Specifically, because they involve non-overlapping subsets of pitches, we can fit independent Bayesian models to obtain posterior samples of $\mathbb{E}[\runs \vert \contextshort, \personelshort, \locationshort, \swing, \outcome]$ and $\P(\outcome \vert \contextshort, \personelshort, \locationshort, \swing)$ for each outcome.
By suitably multiplying and summing these samples according to \Cref{eq:xR}, we obtain posterior samples of $\E[R \vert \contextshort, \personelshort, \locationshort,\swing],$ from which we can immediately compute posterior samples of $\evdiff$ and $\correct.$
Given posterior samples of $\evdiff,$ we use the proportion of samples with positive $\evdiff$ as an estimated probability that swinging is the optimal decision.
We further quantify how much better (or worse) swinging at the pitch is than taking the pitch using the posterior mean and 90\% credible interval (formed using the 5\% and 95\% sample quantiles) of $\evdiff.$

\subsection{Modeling assumptions and fitting}
\label{model_fitting}

\textbf{Expected Runs.} We need to compute run expectancy following the four pitch outcomes shown in \Cref{fig:framework}.
That is, we need to compute $\mathbb{E}[\runs \vert \contextshort, \personelshort, \locationshort, \swing, \outcome]$.
A natural approach begins by first partitioning our dataset of 2,853,912 pitches into bins, one for every combination of variables in $\contextshort,$ $\personelshort,$ $\locationshort,$ $\swing,$ and $\outcome$, and then computing the average number of runs scored subsequently in each bin.
Despite its intuitive appeal, such a binning and averaging procedure is impractical without further simplifying assumptions due to the vast number of combinations.
To wit, in 2019 there were a total of 988 batters, 93 umpires, and 113 catchers.
Accounting for all combinations of just these three aspects of $\personelshort$ requires over 10 million bins, far exceeding the number of pitches in our dataset.

To simplify this estimation, we assume that given the game context, swing decision, and outcome, personnel and pitch location have no predictive effect on expected runs.
Formally, we assume that $\mathbb{E}[\runs \vert \contextshort, \personelshort, \locationshort, \swing, \outcome] = \mathbb{E}[\runs \vert \contextshort, \swing, \outcome].$
While this may seem like a rather strong assumption, we note that it is actually \textit{weaker} than the assumptions underpinning other popular run expectancy models.
%We note that other run expectancy models make similar conditional independent assumptions.
%assumptions about the marginal effects of $\personelshort$ and $\locationshort$ on expected runs.
For example, $\fgre$ assumes that, given the number of outs and the configuration of baserunners, $\runs$ is conditionally independent of all other contextual factors, personnel, and pitch location.
That is, $\fgre$ assumes that $\mathbb{E}[\runs \vert \contextshort, \personelshort, \locationshort, \swing, \outcome] = \mathbb{E}[\runs \vert \texttt{outs}, \texttt{baserunners}]$.
%Our assumption allows for much greater resolution since $\texttt{outs}$ and $\texttt{baserunners}$ are contained $\contextshort$.

Under our assumption, we must now compute $\mathbb{E}[\runs \vert \contextshort, \swing, \outcome].$
We fit a single BART model to estimate the expected runs following a strike, ball, contact, and miss.
Fitting a single model allows us to ``borrow strength'' from related observations with different outcomes.
Such partial pooling may be preferable to separately modeling each outcome since different outcomes can lead to the same game state (e.g.\ swinging strike vs.\ called strike).
We fit our run expectancy model $\bartxr$ using combined pitch data from the 2015 to 2018 MLB seasons, inclusive.

\textbf{Event Probabilities.} We fit BART models with probit links to estimate the strike probability and contact probabilities as functions of $\contextshort$, $\personelshort$, and $\locationshort.$
We fit these models to data from the 2019 MLB season.

%Code to download and pre-process our data, fit the constituent models, and compute the posterior distribution $\evdiff$ and $\correct$ is available at \codelink.

\section{Results}
\label{sec:results}
Code to download and pre-process our data, fit the constituent models, and compute the posterior distributions of $\evdiff$ and $\correct$ is available at \codelink.
All our experiments and analyses were run on a high throughput computing cluster \citep[for details, see][]{chtc}.

\subsection{Model validation}
\label{subsection:model_validation}

We performed several cross-validation studies to understand the predictive accuracy of our BART models for strike probability, contact probability, and run expectancy.

\begin{figure}[ht]
    \centering
    \begin{subfigure}[b]{.45\textwidth}
        \centering
        \includegraphics[width=\textwidth]{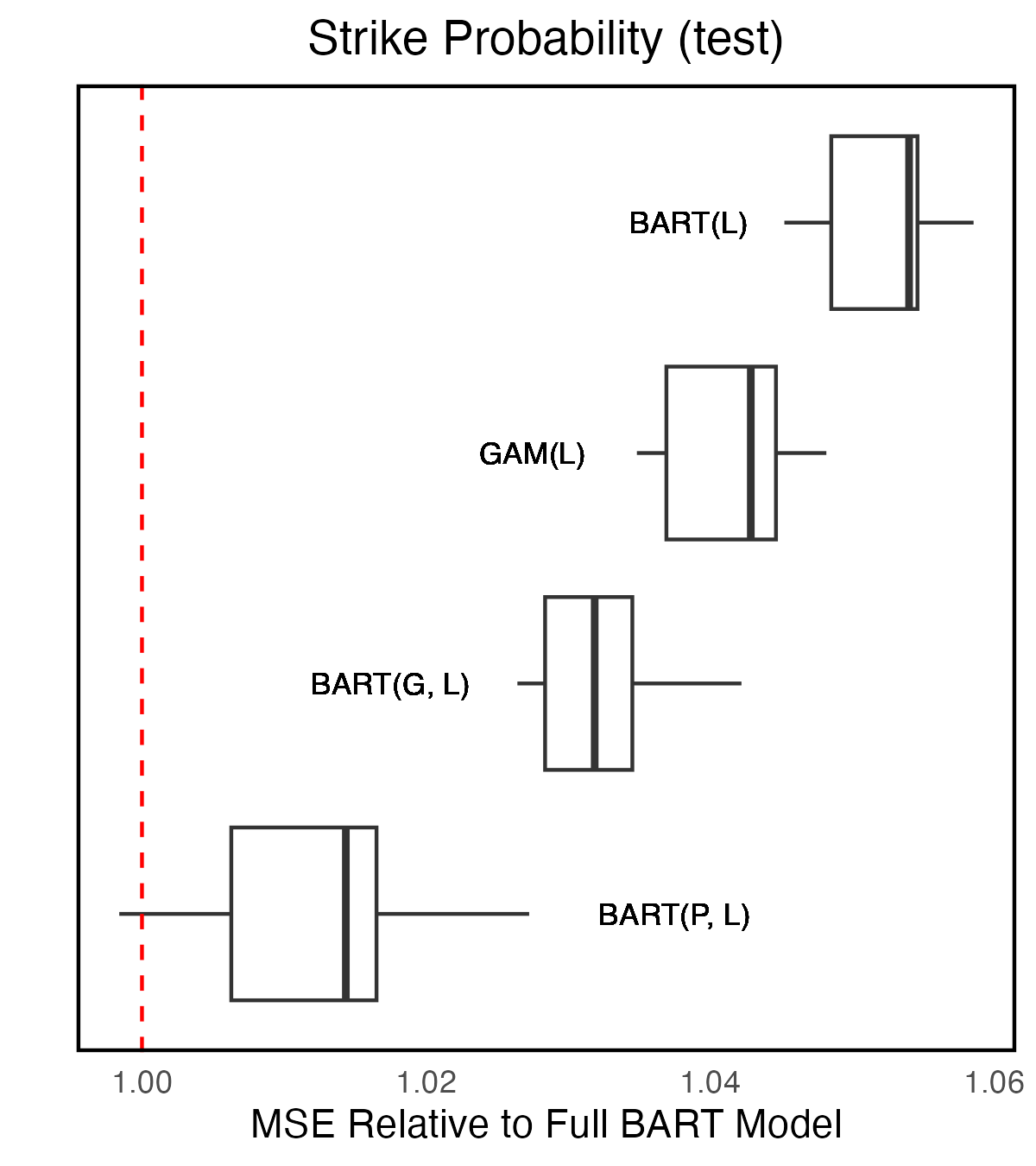}
        \caption{}
        \label{fig:strike_mse}
    \end{subfigure}
    \begin{subfigure}[b]{.45\textwidth}
        \centering
        \includegraphics[width=\textwidth]{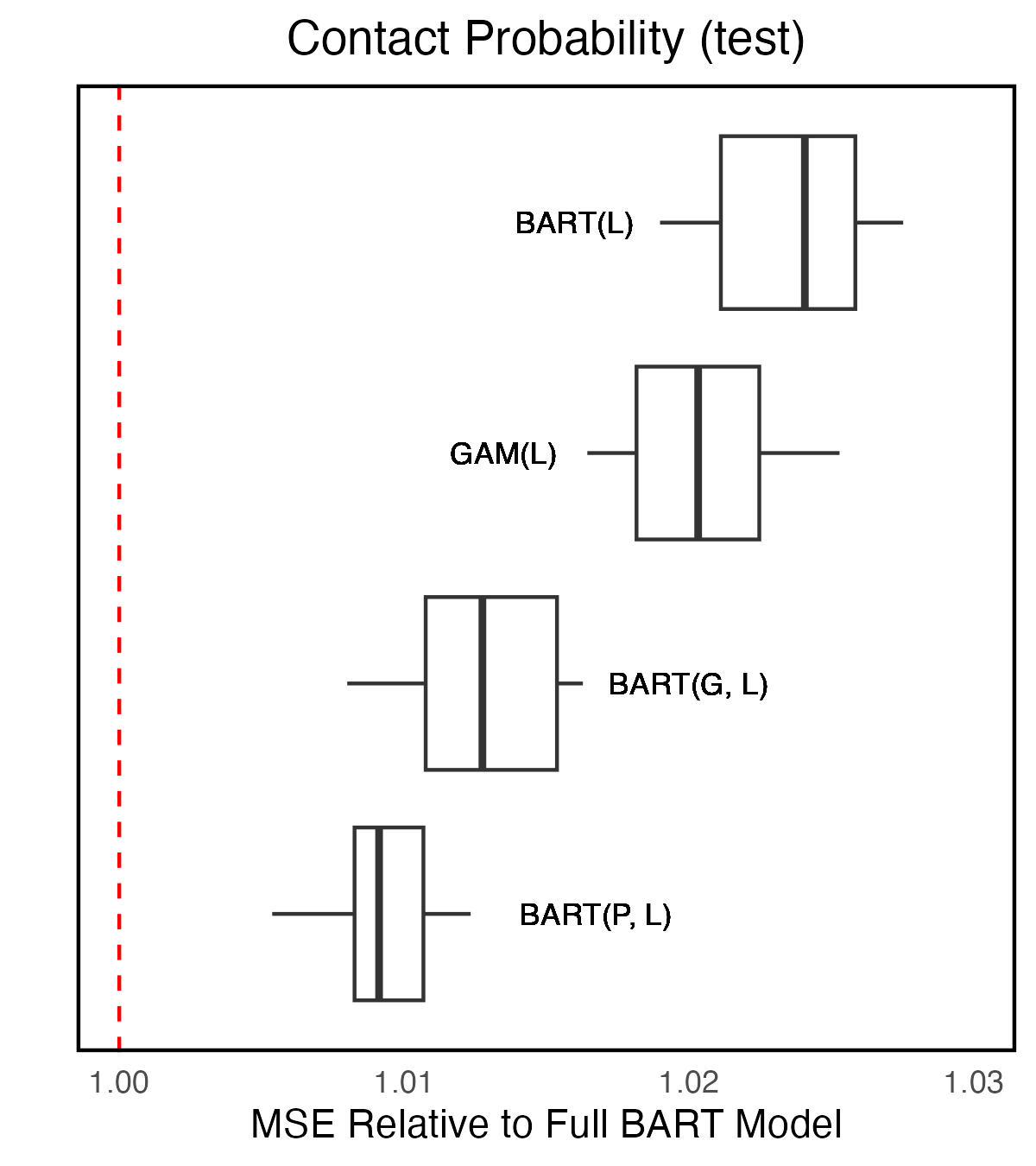}
        \caption{}
        \label{fig:contact_mse}
    \end{subfigure}
    \caption[short]{Out-of-sample mean-squared error relative to a BART model fit with $\contextshort$, $\personelshort$, and $\locationshort$ for strike (a) and contact (b) probabilities.}
    \label{fig:event_prob_mse}
\end{figure}

\textbf{Strike and contact probabilities.}
We compared the out-of-sample performance of our BART models fit with $\contextshort$, $\personelshort$, and $\locationshort$ as predictors for strike and contact probabilities to several competitors including (i) BART models fit with every combination of $\contextshort$, $\personelshort$, and $\locationshort;$ a generalized additive model (GAMs) fit with $\locationshort;$ and the empirical probabilities of each event using 10-fold cross-validation.
We used the \textbf{mgcv} package \citep[version 1.8-41;][]{mgcv_package} to fit GAMs with logit links to predict strike and contact probabilities as a smooth function of location.
We evaluated each model in terms of mean squared error (i.e.\ the Brier score).
\Cref{fig:event_prob_mse} shows the performance of each model relative to BART($\contextshort$, $\personelshort$, $\locationshort$) (i.e.\ the BART model fit with $\contextshort$, $\personelshort$, and $\locationshort$ as predictors).

Unsurprisingly, models using $\locationshort$ as a predictor substantially outperformed models that did not adjust for pitch location.
%We further found that BART achieved slightly smaller average out-of-sample MSE than the other candidate models using location as a predictor.
Interestingly, we found that BART($\personelshort$, $\locationshort$) performed better than BART($\contextshort$, $\locationshort$), but BART($\contextshort$) outperformed BART($\personelshort$).
The fact that these models achieved similar predictive performance, despite using fewer predictors than the full BART model, suggests that location is, by far, the main driver of strike and contact probabilities.
Figures illustrating the relative performance of all other models tested can be found in \Cref{subsection:appendix_cv}.

\textbf{Expected runs.} 
We conducted a comprehensive comparison of $\bartxr$ to 14 variants of $\fgre$.
%We fit each competing model by binning and averaging the number of runs following different game situations \citep{fangraphs_re24}.
We considered every combination of count (12 possibilities), outs (3 possibilities), and baserunners (8 possibilities) as predictors in candidate models for a total of seven unique sets of predictors.
For each set of predictors we fit two types of models: $\rex$ and $\bayesrex$ where \texttt{x} denotes the number of predictors.
$\rex$ models are fit in the same way as $\fgre$.
$\bayesrex$ models are hierarchical Bayesian models of the following form
\begin{align}
\begin{split}
\label{eq:BayesRex}
    \bar{\beta} &\sim \mathcal{N}(0, \tau_{\bar{\beta}}^2) \\
    \tau_{\beta} &\sim \text{half-}t_7 \\
    \sigma^2 &\sim \text{Inverse Gamma}\left(\frac{\nu}{2}, \frac{\nu \lambda}{2}\right) \\
    \beta_{g(i)} &\sim \mathcal{N}(\bar{\beta}, \tau_{\beta}^2) \\
    \tilde{\runs}_{i} &\sim \mathcal{N}(\beta_{g(i)}, \sigma^2)
\end{split}
\end{align}
where $i$ indexes the pitch, $g(i)$ indexes the bin (i.e.,\ combination of count, out, and/or baserunners), $\text{half-}t_{7}$ denotes a $t$-distribution with 7 degrees of freedom truncated to the positive axis, and $\tilde{\runs}$ is $\runs$ standardized to have mean zero and variance one.
We set $\nu$ and $\lambda$ so that the prior probability on the event $\sigma < 1$ is about 90\%.
Our decision to specify the model in \Cref{eq:BayesRex} on the standardized scale and our choices of $\nu$ and $\lambda$ mirror the choices made in default implementations of BART.
We completed our prior specification with the weakly informative choice $\tau_{\bar{\beta}}^2 = 100.$ 
Compared to $\rex,$ which estimates run expectancy in each bin independently, $\bayesrex$ ``borrows strength'' across bins.

\begin{figure}[ht]
    \centering
    \includegraphics[width=0.75\textwidth]{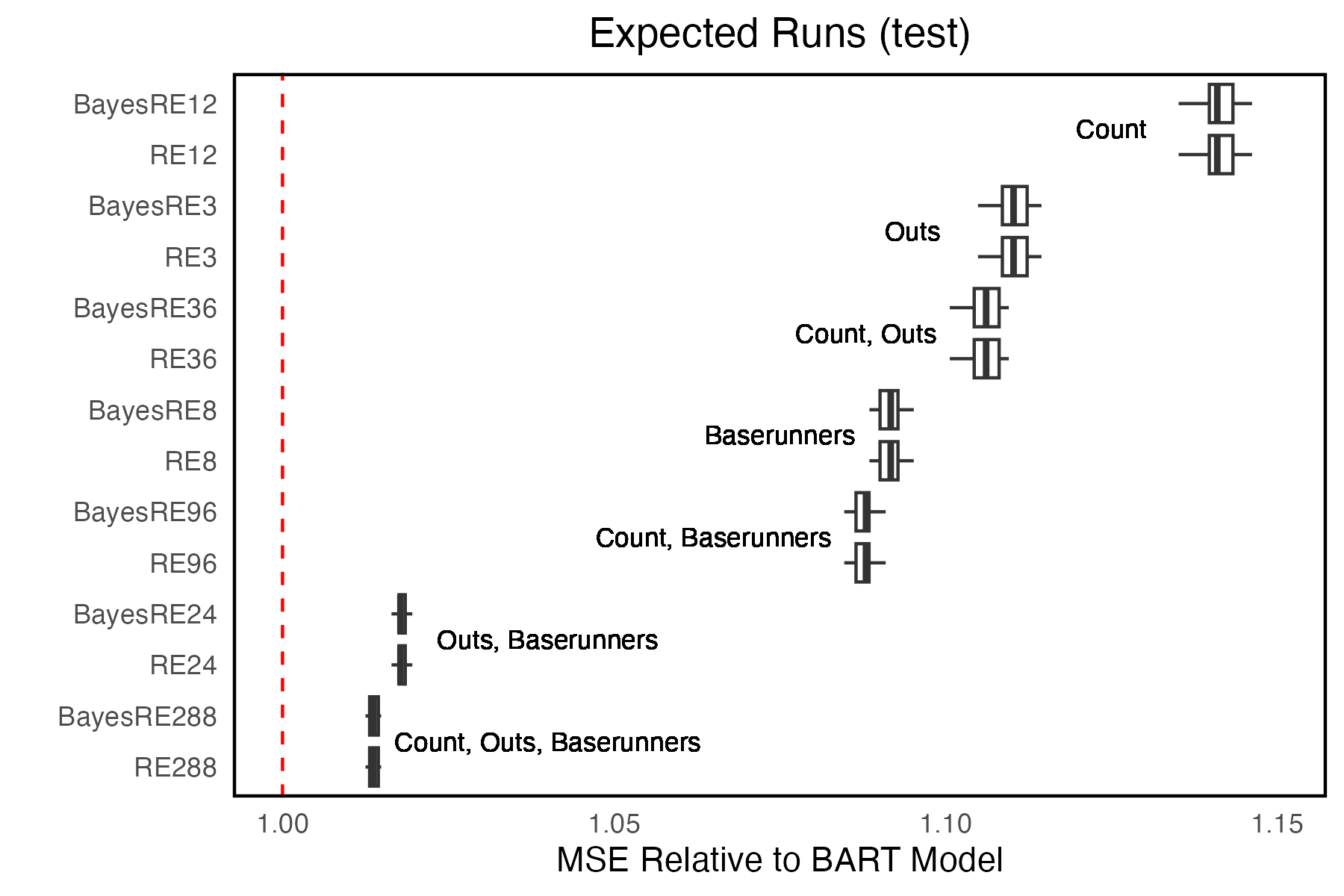}
    \caption[short]{Out-of-sample mean-squared error relative to $\bartxr$ for seven $\rex$ models and seven $\bayesrex$ models.}
    \label{fig:xR_mse}
\end{figure}

\Cref{fig:xR_mse} shows the mean-squared error results of each candidate model relative to $\bartxr$ in 10-fold cross-validation.
We find $\rex$ and $\bayesrex$ models perform similarly for each combination of predictors.
This is perhaps unsurprising given the large number of pitches; basically, the conditional posterior distribution of each $\beta_{g(i)}$ given $\overline{\beta}, \sigma^{2},$ and $\tau_{\beta}$ is sharply concentrated around the average values of the $\tilde{R}_{i}$'s in the bin. 
We find that the models which account for outs and baserunners perform best.
On average, $\bartxr$ was more accurate than \texttt{RE24} by 0.009 runs per pitch and \texttt{RE288} by 0.007 runs per pitch, which respectively correspond to improvements of 1.8\% and 1.4\% in mean square error.
%$\bartxr$ outperforms all other models in both training and testing samples (average RMSE, $\bartxr$: 0.95, \texttt{RE288}: 0.96, \texttt{BayesRE288}: 0.96).

\subsection{Batter evaluation case study}
\label{subsection:case_study}

\begin{figure}[ht]
    \centering
    \begin{subfigure}[b]{0.45\textwidth}
        \centering
        \includegraphics[width=\textwidth]{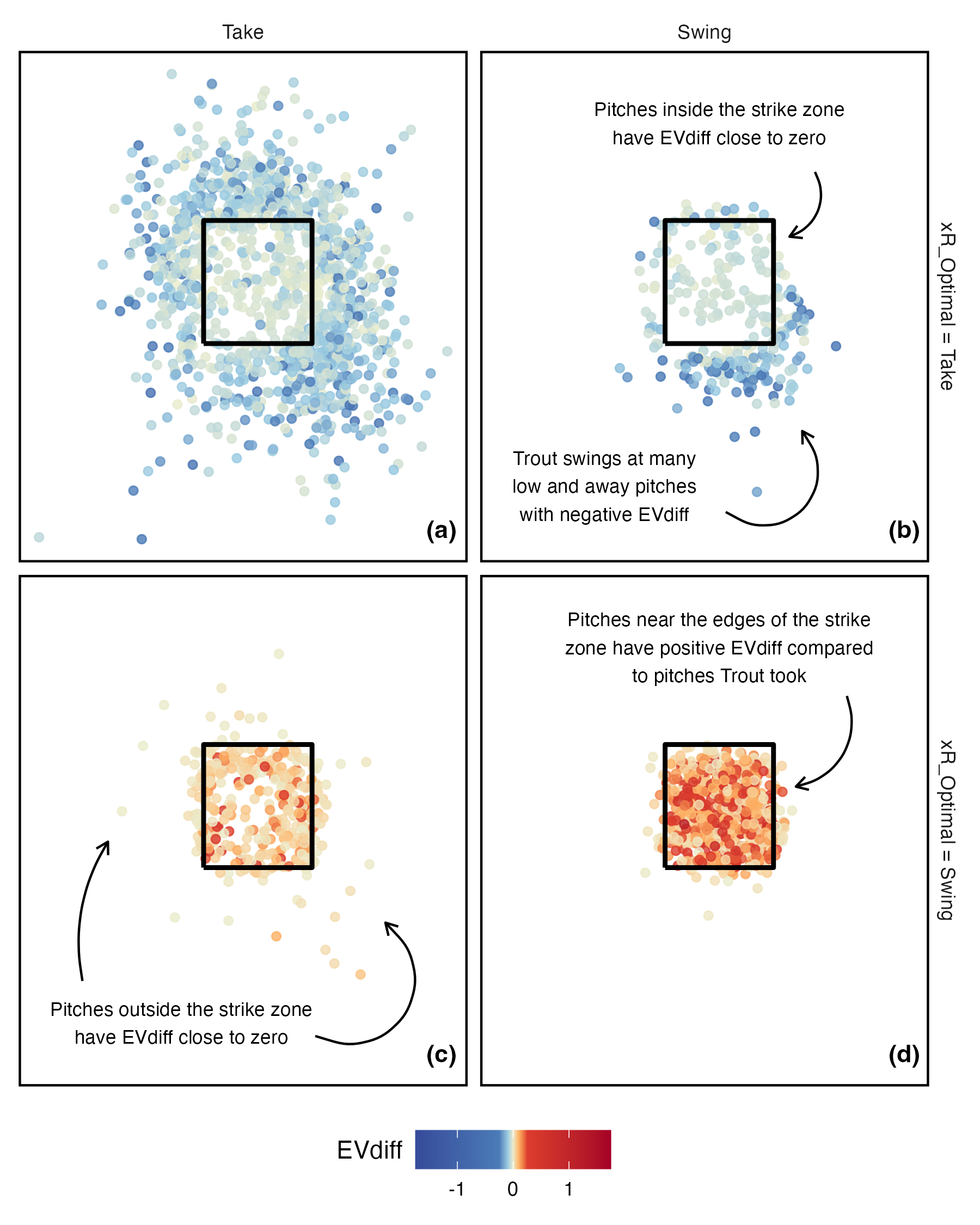}
        \caption{}
        \label{fig:evdiff}
    \end{subfigure}
    \hfill
    \begin{subfigure}[b]{0.45\textwidth}
        \centering
        \includegraphics[width=\textwidth]{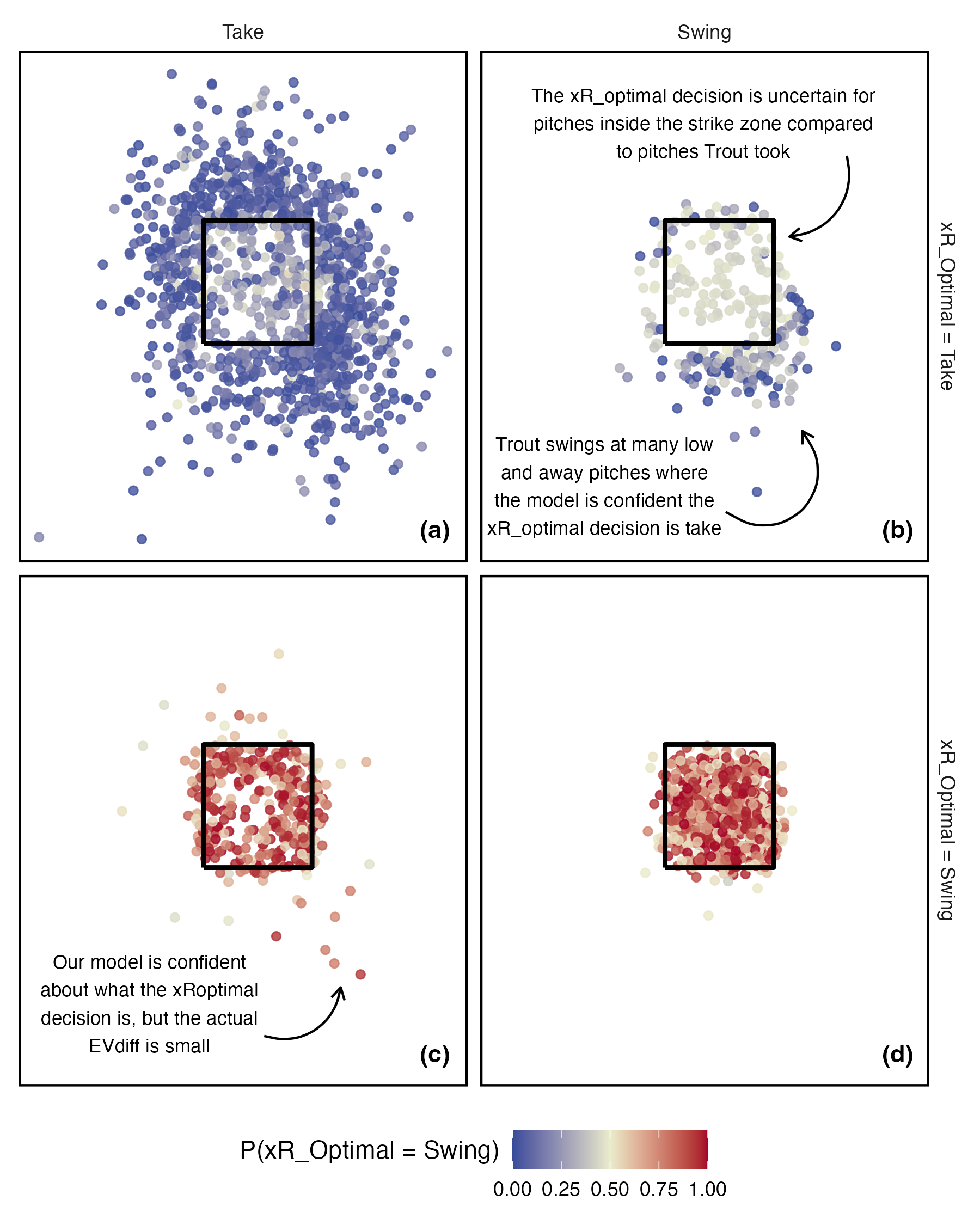}
        \caption{}
        \label{fig:probsgt}
    \end{subfigure}
    \caption{Pitches faced by Mike Trout in the 2019 MLB season where he took (left) and swung (right) and the $\correct$ decision is to take (top) and swing (bottom). Pitches in (a) are shaded based on $\evdiff$ and pitches in (b) are shaded based on $\P(\correct = \swing)$. The darker the shading, the larger the $\evdiff$ (a) and our certainty that the $\correct$ decision is to swing (b). All plots are from the perspective of the home plate umpire.}
    \label{fig:trout_four_panel}
\end{figure}

We now illustrate the type of plate discipline analysis facilitated by our modeling efforts with a case study about Mike Trout in the 2019 MLB season.
%We illustrate a plate discipline analysis for a single batter through a case study of Mike Trout's swing decisions in the 2019 MLB season.
% For each pitch Trout faced in 2019, we calculate the $\evdiff$ by generating posterior samples of our models, predict strike probability, contact probability, and expected runs for each outcome using our sampled models, and computing $\evdiff$ from these predicted quantities.
% Then, we determine the $\correct$ swing decision based on the posterior mean of $\evdiff$ for each pitch Trout faced.
% We can further use these samples to quantify our uncertainty in the optimal decision by computing the proportion of pitches where each decision is $\correct$.
For each pitch Trout faced in 2019, we generated samples of $\evdiff$ and determined the $\correct$ decision based on the posterior mean of $\evdiff$.
We quantified our uncertainty by computing the $\correct$ decision for each individual sample; that is, we computed the proportion of samples in which the $\correct$ decision is to swing.
\Cref{fig:trout_four_panel} shows the results of these calculations for every pitch faced by Trout in the 2019 regular season broken down by the actual decision and the $\correct$ decision.
In \Cref{fig:evdiff}, the darker the shading, the greater the difference in the posterior means of expected runs from swinging and the expected runs from taking.
In \Cref{fig:probsgt}, the darker the shading, the more posterior certainty we have in the $\correct$ decision.

In \Cref{fig:evdiff} and \Cref{fig:probsgt}, the large number of pitches in panels (a) and (d) reveal that Trout's actual decision very often matched our model's $\correct$ decision.
Moreover, these decisions tended to agree with the conventional wisdom that batters should swing at pitches inside the strike zone and take pitches outside the strike zone.
In panels (a) and (d) of the figures, we find, for instance, that Trout tended not to swing at pitches thrown outside the strike zone and rarely took pitches inside the zone.
The dark shading of pitches outside the strike zone in panel (a) illustrate our model's high degree of certainty that the $\correct$ decision is to take the pitch.
Similarly, the dark shading of pitches inside the strike zone in panel (d) illustrate our model's certainty that the $\correct$ decision is to swing.

Perhaps more interesting are those pitches where our model's $\correct$ decision deviated from the conventional wisdom and Trout's actual decision-making (panels (b) and (c) in \Cref{fig:evdiff} and \Cref{fig:probsgt}).
In panel (b), for instance, we see that Trout swung at many pitches inside the strike zone for which our model determined the $\correct$ decision was to take.
As suggested by the relatively light shading of these pitches, we found on further inspection that our model was very uncertain about the $\correct$ decision.
In fact, the posterior distributions of $\evdiff$ for these pitches tended to be nearly symmetric and tightly concentrated around zero. 
Because of the uncertainty in the $\correct$ decision and relatively small magnitude of $\evdiff$ for these pitches, we would not classify Trout's decision to swing at these pitches as bad decisions per se.

In panel (c), however, we find that Trout took several pitches that passed inside the strike zone for which our model determined, with relatively high certainty (as evidenced by the dark shading), that the $\correct$ decision was to swing. 
For these pitches, the posterior distributions of $\evdiff$ were largely concentrated on positive values.
By taking these pitches, our model suggests that Trout cost his team in terms of expected runs. 
Similarly, in panel (b), we find that Trout swung at many low and away pitches that passed outside the strike zone for which our model determined the $\correct$ decision was to take with relatively high certainty.
By swinging at these pitches, our model suggests that Trout cost his team in terms of expected runs. 

Of additional interest are the low-and-away pitches in panel (c) of \Cref{fig:probsgt} where our model is very confident that the $\correct$ decision is to swing.
We found that the posterior mean $\evdiff$ for these pitches is close to zero, suggesting that the cost of making a sub-optimal decision on these pitches is very small.
Nevertheless, it is interesting that our model is so confident that the $\correct$ decision is to swing on these pitches thrown well outside the strike zone, when it is similarly confident that the $\correct$ decision is to take pitches thrown in similar locations in panels (a) and (b) of \Cref{fig:probsgt}.
We speculate that the difference is due, at least in part, to differences in the game contexts in which these pitches were thrown.
%We suspect that

%most other pitches thrown in similar locations the $\correct$ decision is to take.
%Differences in the results of these pitches must be due, at least in part, to differences in game context between these pitches.

\begin{figure}[ht]
    \centering
    \includegraphics[width=0.6\textwidth]{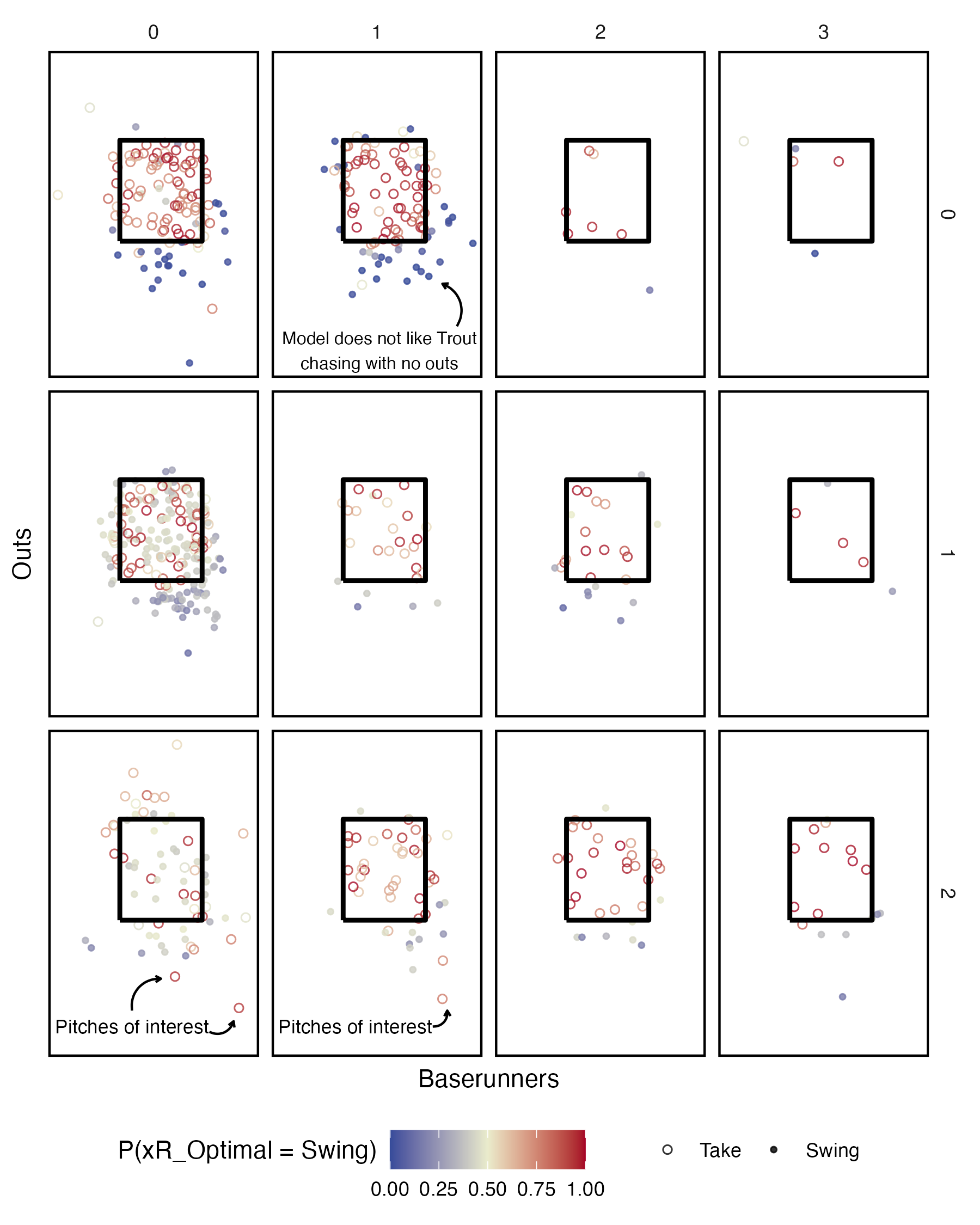}
    \caption[short]{Pitches faced by Mike Trout in the 2019 MLB season where his swing decision conflicts with the $\correct$ decision. Pitches are shaded based on the proportion of samples where the expected runs from swinging is greater than the expected runs from taking.}
    \label{fig:trout_situations}
\end{figure}

% We expect that differences have to do with the context of the game when they were faced.
To probe this possibility, we visualized all pitches faced by Trout broken down by the number of outs and the number of baserunners in \Cref{fig:trout_situations}.
Pitches are colored in \Cref{fig:trout_situations} based on the posterior probability that the $\correct$ decision is to swing.
We see that the low-and-away pitches where the $\correct$ decision is to swing (pitches of interest) occur when there are two outs and few baserunners while the low-and-away pitches where the $\correct$ decision is to take occur when there are no outs.
Such a finding is, we argue, intuitive, because the expected runs from a positive outcome will be similar to the expected runs from a negative outcome when there are two outs and few baserunners.

Finally, although we have focused on Trout in this paper, we can conduct such analyses for any batter using our model results.
We have created an interactive Shiny app \citep[version 1.7.2;][]{rshiny} that performs such visual analysis for all batters who faced at least 1,000 pitches in the 2019 MLB season.
The app is available at \shinylink.

\subsection{Summary metrics}

We can complement our visual analysis using several aggregate metrics to summarize the performance of batters over an entire season.
Importantly, although it is tempting to compare players with these metrics, such comparisons are confounded by differences in pitch situations across players.
Similar to how a batter's RBI depends on how often his teammates get on base, a batter's plate-discipline metrics are dependent on the pitches he sees. 
In \Cref{subsection:limitations} we elaborate on potential methods for making these metrics more comparable and the computational challenges in doing so.

Our first aggregate metric is the proportion of pitches where the batter actually makes the $\correct$ decision.
To account for our uncertainty in the $\correct$ decision, we calculated this proportion for every posterior sample of $\correct,$ and compute a credible interval for this metric.
We found that, on average, batters made the $\correct$ decision 69.4\% of the time in the 2019 MLB season.
For reference, based on the traditional heuristic that a disciplined batter swings at pitches inside the strike zone and takes pitches outside the strike zone, batters made the disciplined decision 68.4\% of the time in the 2019 MLB season.
% source: https://www.fangraphs.com/leaders.aspx?pos=all&stats=bat&lg=all&qual=0&type=5&season=2019&month=1000&season1=2019&ind=0&team=0%2Css&rost=0&age=0&filter=&players=0&startdate=2019-01-01&enddate=2019-12-31&sort=8%2Cd
% calculation: zswing * zone + (1-zone) * (1 -  oswing) = .685 * .418 + (1 - .418) * (1 - .316)
In our dataset, Jonathan Luplow appeared to be the most disciplined batter, making the $\correct$ decision on 76.8\% of pitches (90\% credible interval: [46.1\%, 88.7\%]).
The least disciplined batter in our dataset, Jorge Alfaro, made the $\correct$ decision on 61.4\% of pitches (90\% credible interval: [37.7\%, 83.2\%]).
These numbers closely align with traditional plate discipline metrics that credit Luplow and Alfaro with making the disciplined decision 76.4\% and 60.7\% of the time, respectively.

\begin{figure}[ht]
    \centering
    \includegraphics[width=0.75\textwidth]{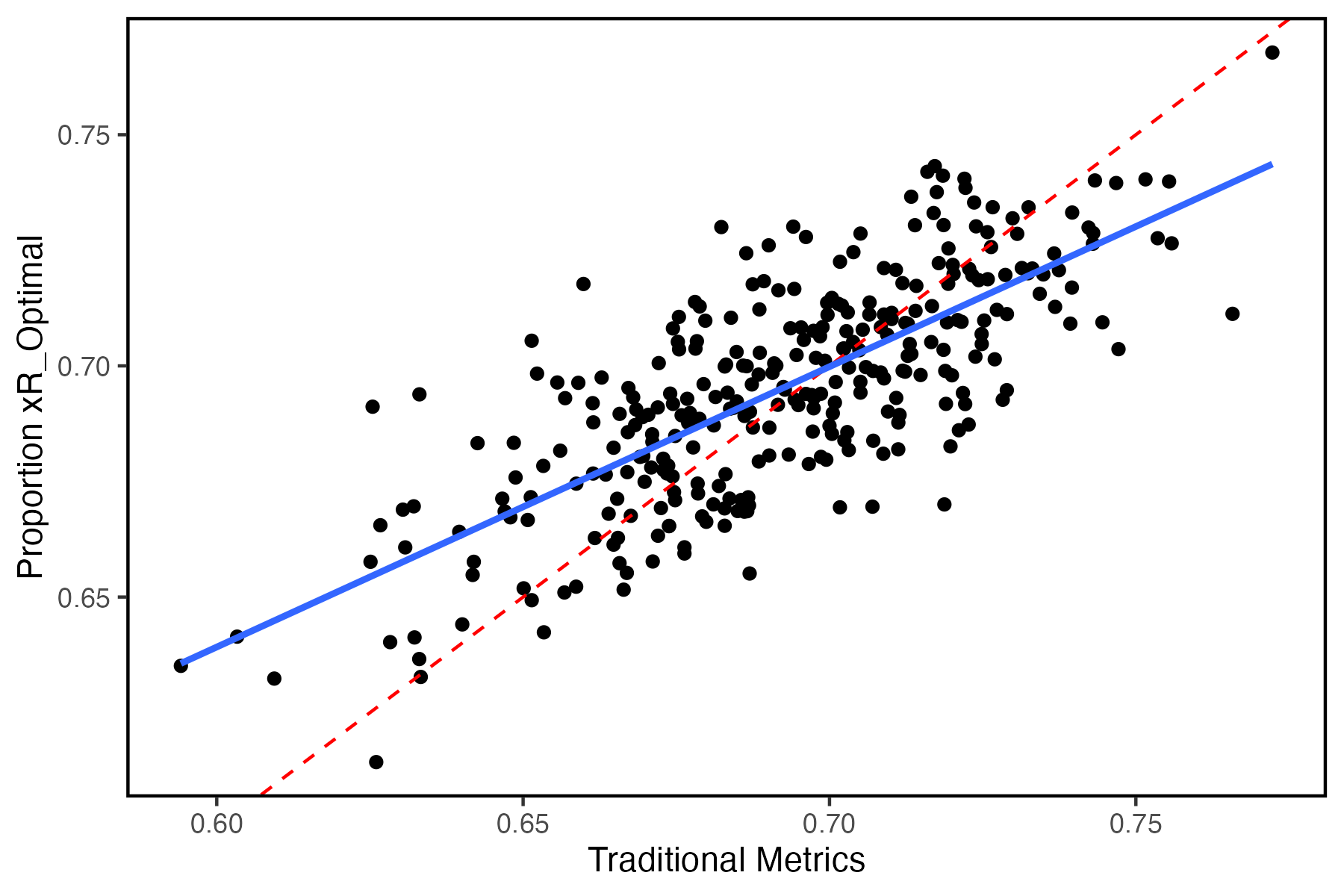}
    \caption[short]{Proportion of pitches where the batter made the $\correct$ decision vs. proportion of pitches where the batter swung (resp. took) pitches inside (resp. outside) the strike zone. Includes all batters who faced at least 1000 pitches in the MLB 2019 regular season. The blue line is the least-squares regression line and the dashed red line is the line the data would follow if both metrics were identical.}
    \label{fig:comparison_2019}
\end{figure}

\Cref{fig:comparison_2019} compares the proportion of $\correct$ decisions to the proportion of traditionally ``correct'' decisions.
We find a moderate correlation (0.750) between the two metrics indicating that batters who follow the traditional plate discipline heuristics tend to make the $\correct$ decision more often.
We observe a shrinkage-like effect where the distribution of batters based on our metric are more tightly clustered around the sample average compared to traditional metrics.
This finding is unsurprising, given the partial pooling performed by our BART models.
%due to the partial-pooling of batters which is a feature of our modeling choice.
We also found that our metric has similar year-to-year consistency as traditional metrics (see \Cref{tab:correlation_tables} in \Cref{subsection:supplementary_tables}).

\begin{figure}[ht]
    \centering
    \begin{subfigure}[b]{0.3\textwidth}
        \centering
        \includegraphics[width=\textwidth]{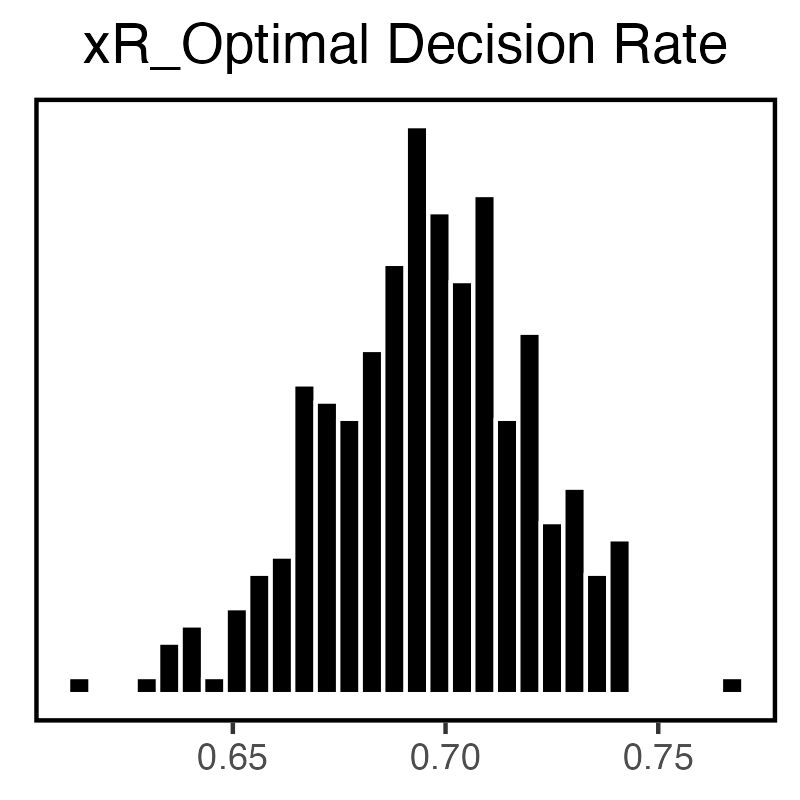}
        % \caption{$\correct$ decision rate}
    \end{subfigure}
    \hfill
    \begin{subfigure}[b]{0.3\textwidth}
        \centering
        \includegraphics[width=\textwidth]{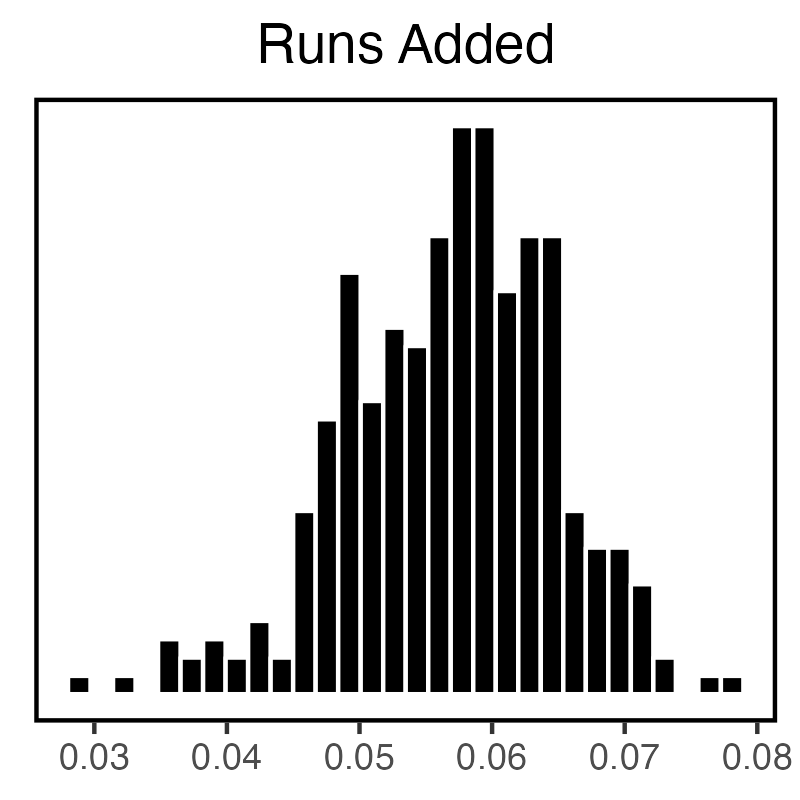}
        % \caption{expected runs added}
    \end{subfigure}
    \hfill
    \begin{subfigure}[b]{0.3\textwidth}
        \centering
        \includegraphics[width=\textwidth]{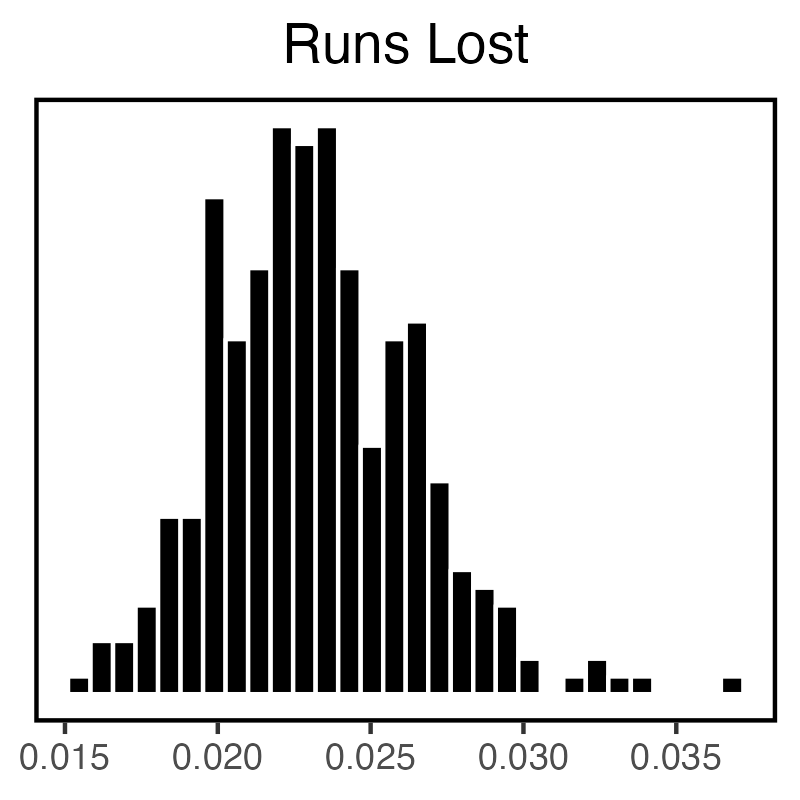}
        % \caption{expected runs lost}
    \end{subfigure}
    \caption{Distributions of posterior means of batters who faced at least 1000 pitches in the 2019 MLB season for each summary metric.}
    \label{fig:metric_dist}
\end{figure}

Our second aggregate metric measures the total impact a batter's decisions have on expected runs over an entire season by taking the difference in expected runs between the decision the batter made and the alternative decision and summing over every pitch a batter faces.
In terms of the plots in \Cref{fig:trout_four_panel}, let $\sum_p \evdiff$ be the sum of $\evdiff$ of all the pitches shown in panel $p$.
Then the expected runs added over an entire season is 
$$
(\sum_d \evdiff - \sum_c \evdiff) - (\sum_a \evdiff - \sum_b \evdiff).
$$
\Cref{fig:metric_dist} shows a histogram of the posterior mean of expected runs added for all batters who faced at least 1000 pitches in the 2019 season.
We find the difference between the best and worst batters according to this metric is not large --- only about 0.1 expected runs.
We further find that there is considerable overlap in the 90\% credible intervals of the expected runs added (see \Cref{fig:added_ci_boxplot}).
For instance, we estimate the top batter, Jordan Luplow, on average adds 0.08 expected runs per pitch due to his decision-making with a 90\% credible interval of [-0.06, 0.21] while the worst batter, Javier Baez, adds 0.03 expected runs per pitch on average with a 90\% credible interval of [-0.13, 0.18].
%These results seem to suggest that there is not a substantial amount of detectable variation in the plate discipline of MLB batters.
% We estimate the top batter, Jordan Luplow, on average adds 0.0778 expected runs per pitch due to his decision-making with a 90\% credible interval of [-0.0630, 0.2117] while the worst batter, Javier Baez, adds 0.0286 expected runs per pitch on average with a 90\% credible interval of [-0.1254, 0.1790].

While informative, the expected runs added metric has limitations.
We argue that if a batter faces many low-leverage pitches (i.e.\ pitches where $\lvert\evdiff\rvert$ is close to zero) he will have fewer opportunities to pick up added runs than batter's facing many high-leverage pitches (i.e.\ pitches where $\lvert\evdiff\rvert$ is large).
To account for this situation, we calculated runs lost as the minimum of 0 and the runs added from a pitch.
To motivate this metric, we argue that when a batter faces a low-leverage pitch, his decision does not matter since it may be unclear what the best decision is.
For example, in \Cref{fig:trout_four_panel} panel (b) there are many pitches inside the strike zone that Trout swings at that, while the $\correct$ decision is to take, there is still considerable uncertainty in the $\correct$ decision.
Since the most a batter can be punished on a given pitch is the expected runs lost by not making the $\correct$ decision, on low-leverage pitches the batter will either not be penalized or be penalized by a small amount.
When a batter faces a high-leverage pitch, the $\correct$ decision should be obvious, so a batter that does not make the $\correct$ decision on these pitches will be severely punished.
In terms of the plots in \Cref{fig:trout_four_panel}, runs lost is
$$
\sum_c \evdiff - \sum_b \evdiff.
$$

Results of expected runs lost are similar to the results of expected runs added: the absolute differences between the best and worst batters are small (see \Cref{fig:metric_dist}) with considerable overlap in credible intervals (see \Cref{fig:lost_ci_boxplot}).
We estimate the top batter, Andrew McCutchen, has an average loss of 0.03 expected runs per pitch with a 90\% credible interval of [0.01, 0.09] while the worst batter, Jorge Alfaro, lost 0.07 expected runs per pitch on average with a 90\% credible interval of [0.01, 0.20].
% We estimate the top batter, Andrew McCutchen, has an average loss of 0.03406 expected runs per pitch with a 90\% credible interval of [0.00979, 0.09014] while the worst batter, Jorge Alfaro, lost 0.06647 expected runs per pitch on average with a 90\% credible interval of [0.01376, 0.19794].

For all three of these metrics, we find that there are a few batters who do much better than everyone else and a few who are much worse.
\Cref{fig:metric_dist} shows the distribution of these metrics for batters that faced at least 1000 pitches in the 2019 MLB season.

%While it is tempting to use these metrics to compare the plate discipline of different batters, note that these metrics are confounded by the location of pitches a batter faces, so they are not necessarily directly comparable across batters.
%We will return to this point in \Cref{sec:discussion}.
%Marginalizing out the effects of location may seem sensible; but, we do not think this is how batter's make decisions --- we suspect batter's do not make swing decisions prior to a pitch being thrown.
%Because of this and the high computational costs of conducting such an analysis we have decided not to do this.

\section{Discussion}
\label{sec:discussion}
We developed a three-step framework for estimating the optimal swing/take decision of batters in Major League Baseball.
In the first step of our framework, we estimate, for a given pitch, (i) the probability the umpire will call a strike; (ii) the probability a batter will make contact; and (iii) the expected number of runs the batting team will score in the remainder of the inning after a ball, strike, contact, and miss.
Then, we combine these estimates to calculate the expected runs following a swing and a take decision.
Finally, we determine the $\correct$ decision to be the one that leads to more expected runs.
We adopt a Bayesian approach which allows us to propagate uncertainty from each intermediate estimate to our evaluation of the $\correct$ decision.

Our findings have several implications for Major League Baseball teams.
First, we can determine those batters who consistently make $\correct$ swing decisions, which can aid in player evaluation.
We can also identify pitch locations where a batter consistently makes suboptimal swing decisions.
Batters can use this information to make adjustments to their decision-making, while pitchers can use this information to target locations where a batter is likely to make a decision that costs his team runs.
Finally, we can identify locations where pitches are likely to lead to a significant increase in expected runs (i.e.\ locations with high $\evdiff$).
Pitchers can use this information to avoid these locations to minimize the opportunities for the batting team to score.

We found that across the league, batters made the $\correct$ decision 69.4\% of the time according to our model, which aligns with the findings of other plate discipline studies. 
Critically, we should ask ourselves why are our results so similar to the simpler traditional plate discipline metrics.
One explanation is that, despite their somewhat heavy-handed assumptions (e.g.\ one should swing at every pitch in the strike zone and take every pitch outside the zone), traditional metrics offer a reasonably accurate approximation of our more fine-grained model.

Like umpires, batters are not robots: just as umpires do not strictly adhere to the rulebook definition of the strike zone, batters almost certainly consider more than just expected runs when making swing decisions.
The fact that batters deviated from the $\correct$ decision about 30\% of the time suggests that batters do not always consider expected runs when making swing decisions.
For example, a player chasing a home run record (e.g.,\ Aaron Judge at the end of the 2022 season) may make swing decisions to maximize expected home runs.
While these decisions are not $\correct$, they might be optimal when viewed through the lens of expected home runs.

\begin{comment}
As an extreme example, consider Aaron Judge chasing the AL home run record towards the end of the 2022 MLB season.
During the last few weeks of the season, Judge was swinging at a lot of pitches to make a final push at the record.
While many would agree this is generally not the best policy, is it fair to say Judge's decisions were ``undisciplined''?
What if Judge's manager had given him a green light to swing for the fences in these final weeks?
A case could be made that getting the home run record is a more meaningful achievement than winning an extra game or two at the end of a long season.
More routine examples where batters may make decisions conflicting with the $\correct$ decision could include situations where your team's best batter is on-deck, or you are near the top of the line-up and don't want to risk hitting into a double-play.

The answer may be that it is impossible to pick a single criterion on which to evaluate a batter's swing decisions, but that doesn't mean we shouldn't do anything.
We choose to use expected runs as our measure to compare the results of swing decisions following a sensible argument that a batter should make the swing decision that leads to more runs for his team.
Using expected runs also yielded interpretable results that can be easily understood.
However, our modular framework gives us the flexibility to choose any other metric to evaluate batters such as change in win probability, expected home runs, or expected wOBA added.
\end{comment}

\subsection{Limitations and future work}
\label{subsection:limitations}

We presented an analysis of plate discipline that asserts the ``disciplined'' swing decision is the one that leads to a greater number of expected runs.
While we believe our assertion is reasonable, an honest argument could be made that this is an incorrect characterization of a ``disciplined'' swing decision.
In such cases, our modular framework gives us the flexibility to use any other objective to evaluate batters.
For example, we could replace $\runs$ in each branch of \Cref{fig:framework} with another team outcome (e.g.,~win probability) or individual outcome (e.g.,~home runs, wOBA, OPS) and follow the same strategy: estimate intermediate probabilities and conditional expectations; compute the expected outcome following a swing and following a take; and determine the optimal decision.

While we used BART to fit the three models in the first step of our framework, other choices are possible.
Indeed, we selected BART for its ease-of-use: we did not have to manually pre-specify the functional forms of the called strike and contact probabilities and run expectancy, which we suspected depended on complicated non-linearities and interactions.
Although our BART-based models slightly outperformed parametric alternatives, these differences in out-of-sample predictive power were not especially large.
For instance, we found that our BART model of called strike and contact probabilities, which accounted for location, game state, and pitch personnel, achieved very similar predictive performance as a generalized additive model that only accounted for location.
Such a finding, to us, suggests that pitch location is, by far, the main driver of strike and contact probabilities, with other predictors like outs or count or player identifies contributing little additional predictive power.
We also found that the difference in out-of-sample RMSE between our BART-based run expectancy model, $\bartxr,$ and the best-performing alternative based on binning and averaging was about 0.01 runs.
While this is a small difference on a per-pitch basis, these differences can magnify over the course of an entire season.

More substantively, we only considered two possible outcomes following a swing decision, a miss or contact.
We could extend our framework to account for more post-swing outcomes like miss, foul, out, single, double, triple, and home run in one of two ways.
First, we could replace our binary contact probability model with \citet{Murray2021}'s multinomial logistic BART model.
In the context of \Cref{fig:framework}, this would involve replacing the \texttt{contact} branch with several branches.
Alternatively, we could augment our existing framework with a further model of these outcomes \textit{conditional} on making contact.
That is, we could add additional child branches to the \texttt{contact} branch of \Cref{fig:framework}, one for each potential outcome following contact.
We note that the developers of EAGLE pursued the first strategy.
We suspect, however, that the second approach would lead to less uncertainty about the $\correct$ decision as the overall accuracy of our contact/miss model is better than the reported accuracy of EAGLE's multi-outcome model. 

Beyond a pitch-by-pitch visual assessment, we introduced three aggregate metrics that tracked the proportion of times batters made the $\correct$ decision and the expected number of runs added or lost due to plate discipline across an entire season.
Although it is tempting to compare these metrics across players, these metrics are confounded by pitch location and the context in which players see different pitches.
Simply put, because the distribution of pitches one batter sees may differ from the distribution seen by another batter, it is difficult to directly compare their aggregated plate discipline metrics.
To overcome such confounding, we could first marginalize $\evdiff$ over $\locationshort$ and $\contextshort$ before computing in a manner similar to \citet{safe}'s spatially aggregate fielding evaluation. 
While such spatially and contextually aggregated plate discipline metrics are intuitively appealing, computing them introduces considerable computational challenges.
Basically, to marginalize over $\locationshort$ and $\contextshort,$ we must make predictions for every combination of batter and pitch in our dataset. 
We leave efficient computation of the approximately seven hundred million combinations required for such marginal combinations to future work.

Although we did not control for pitch-related factors like type, velocity, or spin, including them in our models is relatively straightforward.
Accounting for pitch sequencing is somewhat harder.
One approach would be to include information about the previous $k$ pitches as additional covariates.
However, determining the appropriate lag $k$ is highly non-trivial, since it may vary batter-to-batter (in the case of contact probability) or umpire-to-umpire (in the case of called strike probability).
Relatedly, our strike and contact probability models both condition on the location of the pitch as it crosses the front edge of home plate.
Insofar as batters decide to swing or take the pitch before it reaches home plate, one could reasonably argue that such precise location information is unavailable to the batter.
Rather than omit pitch location from our strike and contact probability models, a more realistic model would incorporate the trajectory (or portions thereof) of the pitch as it travels from the pitcher's hand to home plate.
Incorporating \textit{functional} predictors into BART and other tree-based models is an interesting and important methodological challenge.

{
\small
\singlespacing
\bibliographystyle{apalike}
\bibliography{plate_discipline_refs}

\begin{thebibliography}{}

\bibitem[Arthur, 2014a]{moonshot-pd}
Arthur, R. (2014a).
\newblock Moonshot: The new best way to measure plate discipline.
\newblock
  \url{https://www.baseballprospectus.com/news/article/25008/moonshot-the-new-best-way-to-measure-plate-discipline/}.

\bibitem[Arthur, 2014b]{moonshot-sz}
Arthur, R. (2014b).
\newblock Moonshot: The victims of a bad strike zone.
\newblock
  \url{https://www.baseballprospectus.com/news/article/24862/moonshot-the-victims-of-a-bad-strike-zone/}.

\bibitem[Brill et~al., 2023]{Brill2022-tp}
Brill, R.~S., Deshpande, S.~K., and Wyner, A.~J. (2023).
\newblock A {Bayesian} analysis of the time through the order penalty in
  baseball.
\newblock {\em Journal of Quantitative Analysis in Sports}.

\bibitem[{Center for High Throughput Computing}, 2006]{chtc}
{Center for High Throughput Computing} (2006).
\newblock Center for high throughput computing.

\bibitem[Chang et~al., 2022]{rshiny}
Chang, W., Cheng, J., Allaire, J., Sievert, C., Schloerke, B., Xie, Y., Allen,
  J., McPherson, J., Dipert, A., and Borges, B. (2022).
\newblock {\em {shiny}: Web application framework for R}.

\bibitem[Chen et~al., 2016]{Chen2016}
Chen, D.~L., Moskowitz, T.~J., and Shue, K. (2016).
\newblock Decision mkaing under the {Gambler's Fallacy}: Evidence from asylum
  judges, loan officers, and baseball umpires.
\newblock {\em The Quarterly Journal of Economics}, 131(3):1181--1242.

\bibitem[Chipman et~al., 2010]{Chipman2010}
Chipman, H.~A., George, E.~I., and McCulloch, R.~E. (2010).
\newblock {BART}: Bayesian additive regression trees.
\newblock {\em Annals of Applied Statistics}, 4(1):266--298.

\bibitem[Deshpande, 2023]{Deshpande2022}
Deshpande, S.~K. (2023).
\newblock {flexBART}: Flexible {Bayesian} regression trees with categorical
  predictors.
\newblock \textit{arXiv preprint arXiv:2211.04459}.

\bibitem[Deshpande and Wyner, 2017]{DeshpandeWyner2017}
Deshpande, S.~K. and Wyner, A.~J. (2017).
\newblock A hierarchical {Bayesian} model of pitch framing.
\newblock {\em Journal of Quantitative Analysis in Sports}, 13(3):95--112.

\bibitem[Green and Daniels, 2022]{GreenDaniels2021_instinct}
Green, E. and Daniels, D. (2022).
\newblock Bayesian instinct.
\newblock \textit{SSRN preprint 2916929}.

\bibitem[Green and Daniels, 2014]{Green2014-pj}
Green, E. and Daniels, D.~P. (2014).
\newblock What does it take to call a strike? {Three} biases in umpire decision
  making.
\newblock In {\em MIT Sloan Sports Analytics Conference}.

\bibitem[Jensen et~al., 2009]{safe}
Jensen, S.~T., Shirley, K.~E., and Wyner, A.~J. (2009).
\newblock Bayesball: A {Bayesian} hierarchical model for evaluating fielding in
  {Major League Baseball}.
\newblock {\em Annals of Applied Statistics}, 3(2):491--520.

\bibitem[Kim and King, 2014]{KimKing2014}
Kim, J.~W. and King, B.~G. (2014).
\newblock Seeing stars: {Matthew} effects and status bias in {Major League
  Baseball} umpiring.
\newblock {\em Management Science}, 60(11):2619--2644.

\bibitem[Lindbergh, 2013]{lindbergh-2013}
Lindbergh, B. (2013).
\newblock The art of pitch framing.
\newblock
  \url{http://grantland.com/features/studying-art-pitch-framing-catchers-such-francisco-cervelli-chris-stewart-jose-molina-others/}.

\bibitem[Marchi, 2011]{Marchi-2011}
Marchi, M. (2011).
\newblock Evaluating catchers: Quantifying the framing pitches skill.
\newblock
  \url{https://tht.fangraphs.com/evaluating-catchers-quantifying-the-framing-pitches-skill/}.

\bibitem[Mills, 2014]{Mills2014-hd}
Mills, B.~M. (2014).
\newblock Social pressure at the plate: Inequality aversion, status, and mere
  exposure.
\newblock {\em Managerial and Decision Economics}, 35(6):387--403.

\bibitem[Mills, 2017]{Mills2017}
Mills, B.~M. (2017).
\newblock Technological innovations in monitoring and evaluation: evidence of
  performance impacts among {Major League Baseball} umpires.
\newblock {\em Labour Economics}, 46:189--199.

\bibitem[Mould and Anderson, 2022a]{eagel-p1}
Mould, J. and Anderson, D. (2022a).
\newblock Quantifying hitter plate discipline with eagle: Part 1.
\newblock
  \url{https://www.baseballprospectus.com/news/article/74173/quantifying-hitter-plate-discipline-with-eagle-part-1/}.

\bibitem[Mould and Anderson, 2022b]{eagel-p2}
Mould, J. and Anderson, D. (2022b).
\newblock Quantifying hitter plate discipline with eagle: Part 2.
\newblock
  \url{https://www.baseballprospectus.com/news/article/74214/quantifying-hitter-plate-discipline-with-eagle-part-2/}.

\bibitem[Murray, 2021]{Murray2021}
Murray, J.~S. (2021).
\newblock Log-linear {Bayesian Additive Regression Trees} for multinomial
  logistic and count regression.
\newblock {\em Journal of the American Statistical Association},
  116(534):756--769.

\bibitem[Petti and Gilani, 2022]{baseballr_package}
Petti, B. and Gilani, S. (2022).
\newblock {\em {baseballr}: Acquiring and analyzing baseball data}.

\bibitem[Slowinski, 2010]{fangraphs_PD_metrics}
Slowinski, P. (2010).
\newblock {Plate Discipline}.
\newblock \url{https://library.fangraphs.com/offense/plate-discipline/}.

\bibitem[Tango et~al., 2007]{tango2007book}
Tango, T., Lichtman, M., and Dolphin, A. (2007).
\newblock {\em The Book: Playing the Percentages in Baseball}.
\newblock Potomac Books.

\bibitem[Vock and Vock, 2018]{VockVock2018}
Vock, D.~M. and Vock, L. F.~B. (2018).
\newblock Estimating the effect of plate discipine using a causal inference
  framework: an application of the {G-computation} algorithm.
\newblock {\em Journal of Quantitative Analysis in Sports}, 14(2):37--66.

\bibitem[Wood, 2017]{mgcv_package}
Wood, S.~N. (2017).
\newblock {\em Generalized Additive Models: An Introduction with {R}}.
\newblock Chapman and Hall/CRC, 2 edition.

\end{thebibliography}
}
\onehalfspacing
\section{Appendix}
\label{sec:appendix}

\subsection{Cross-validation results}
\label{subsection:appendix_cv}

\begin{figure}[ht]
    \centering
    \begin{subfigure}[b]{.45\textwidth}
        \centering
        \includegraphics[width=\textwidth]{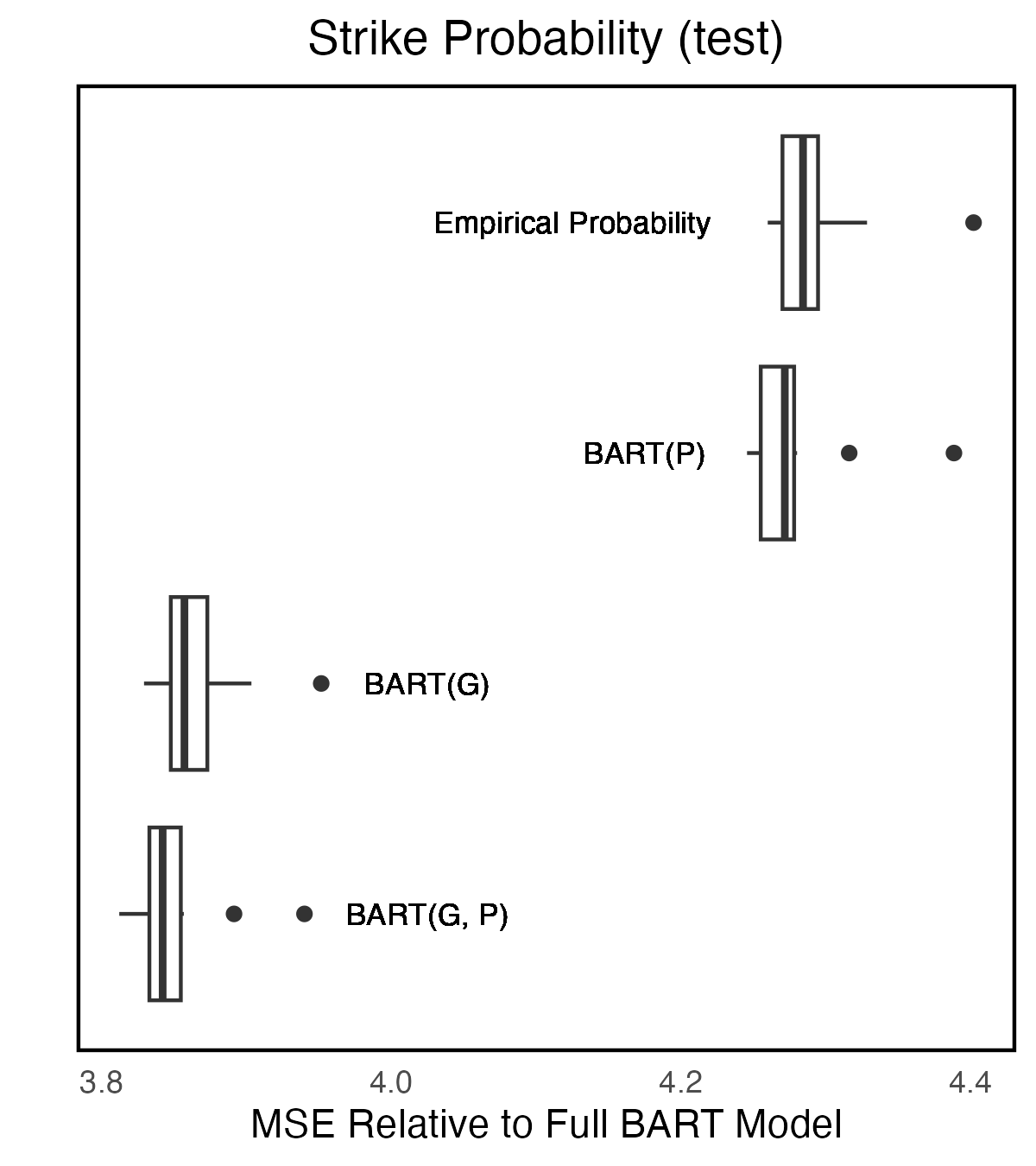}
        \caption{}
        \label{fig:strike_mse_all}
    \end{subfigure}
    \begin{subfigure}[b]{.45\textwidth}
        \centering
        \includegraphics[width=\textwidth]{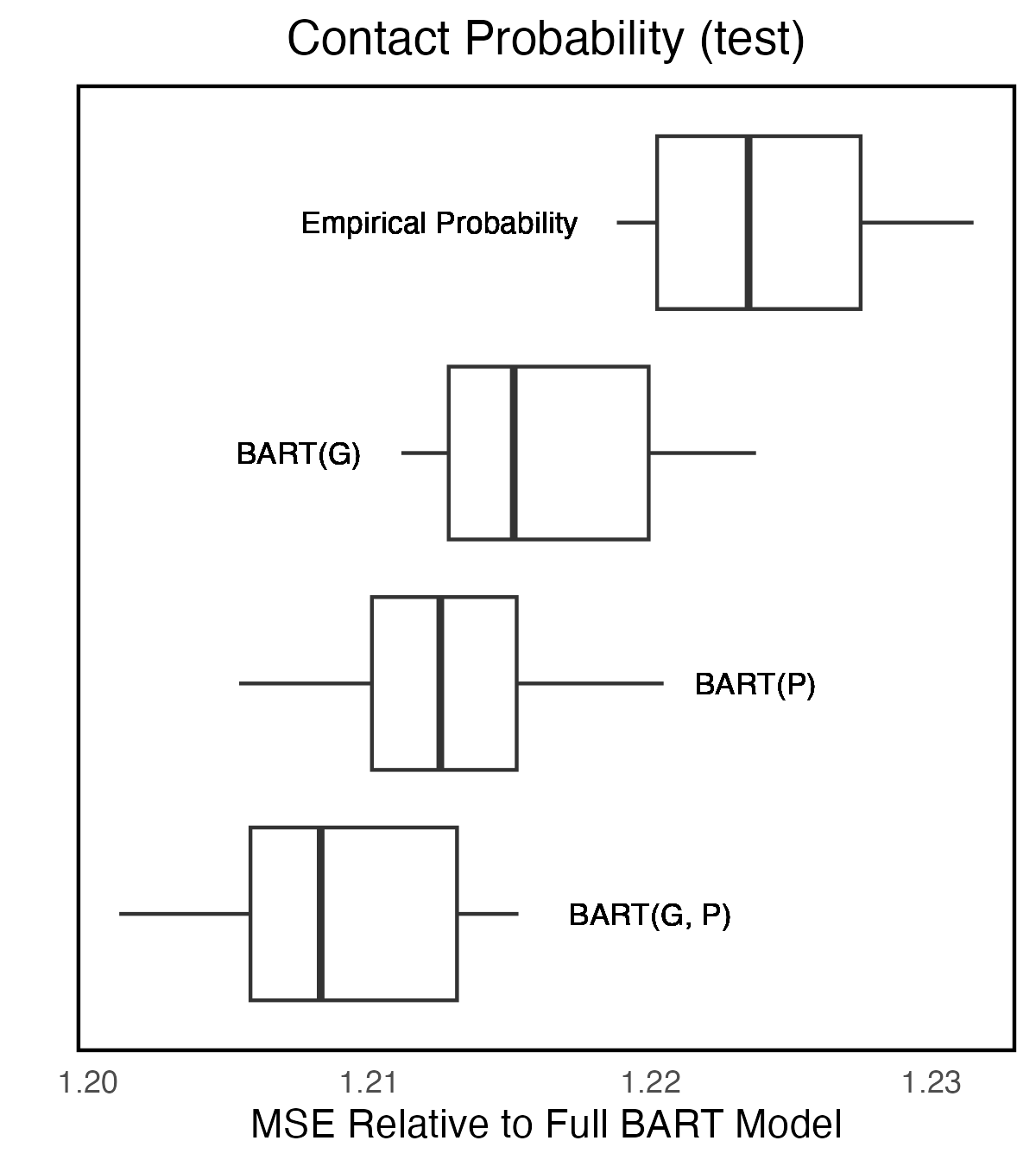}
        \caption{}
        \label{fig:contact_mse_all}
    \end{subfigure}
    \caption[short]{Out-of-sample mean-squared error relative to a BART model fit with $\contextshort$, $\personelshort$, and $\locationshort$ for strike (a) and contact (b) probabilities.}
    \label{fig:event_prob_mse_all}
\end{figure}

\Cref{fig:event_prob_mse_all} reports the relative mean-square error (i.e.\ Brier score) for strike and contact probability BART models fit with every combination of $\contextshort$ and $\personelshort$ relative to the full BART model fit with $\contextshort$, $\personelshort$ and $\locationshort$.
Because these models do not adjust for pitch location $\locationshort,$ the relative MSEs are much larger than those shown in \Cref{fig:event_prob_mse}.

\subsection{Stability of Summary Metrics}
\label{subsection:supplementary_tables}

\Cref{tab:correlation_tables} shows the year-to-year correlation of (i) the proportion of pitches where a batters makes the $\correct$ decision and (ii) the proportion of pitches where a batter swung if the pitch was inside the strike zone and took if the pitch was outside the strike zone.
Batters who faced at least 1000 pitches in all three MLB regular seasons from 2017-2019 were included.

\begin{table}[ht]
    \caption{Year-to-year correlation of metrics.}
    \label{tab:correlation_tables}
    \centering
    \begin{tabular}{c|ccc|ccc}
    & \multicolumn{3}{c|}{Proportion $\correct$} & \multicolumn{3}{c}{Traditional Metrics} \\
    & 2017 & 2018 & 2019 & 2017 & 2018 & 2019 \\
    \hline
    2017 & 1.000 & 0.714 & 0.722 & 1.000 & 0.772 & 0.740 \\
    2018 & 0.714 & 1.000 & 0.735 & 0.772 & 1.000 & 0.790 \\
    2019 & 0.722 & 0.735 & 1.000 & 0.740 & 0.790 & 1.000 \\ \hline
    \end{tabular}
\end{table}

\subsection{Additional figures}
\label{subsection:appendix_figs}

\begin{figure}[ht]
    \centering
    \includegraphics[width=.6\textwidth]{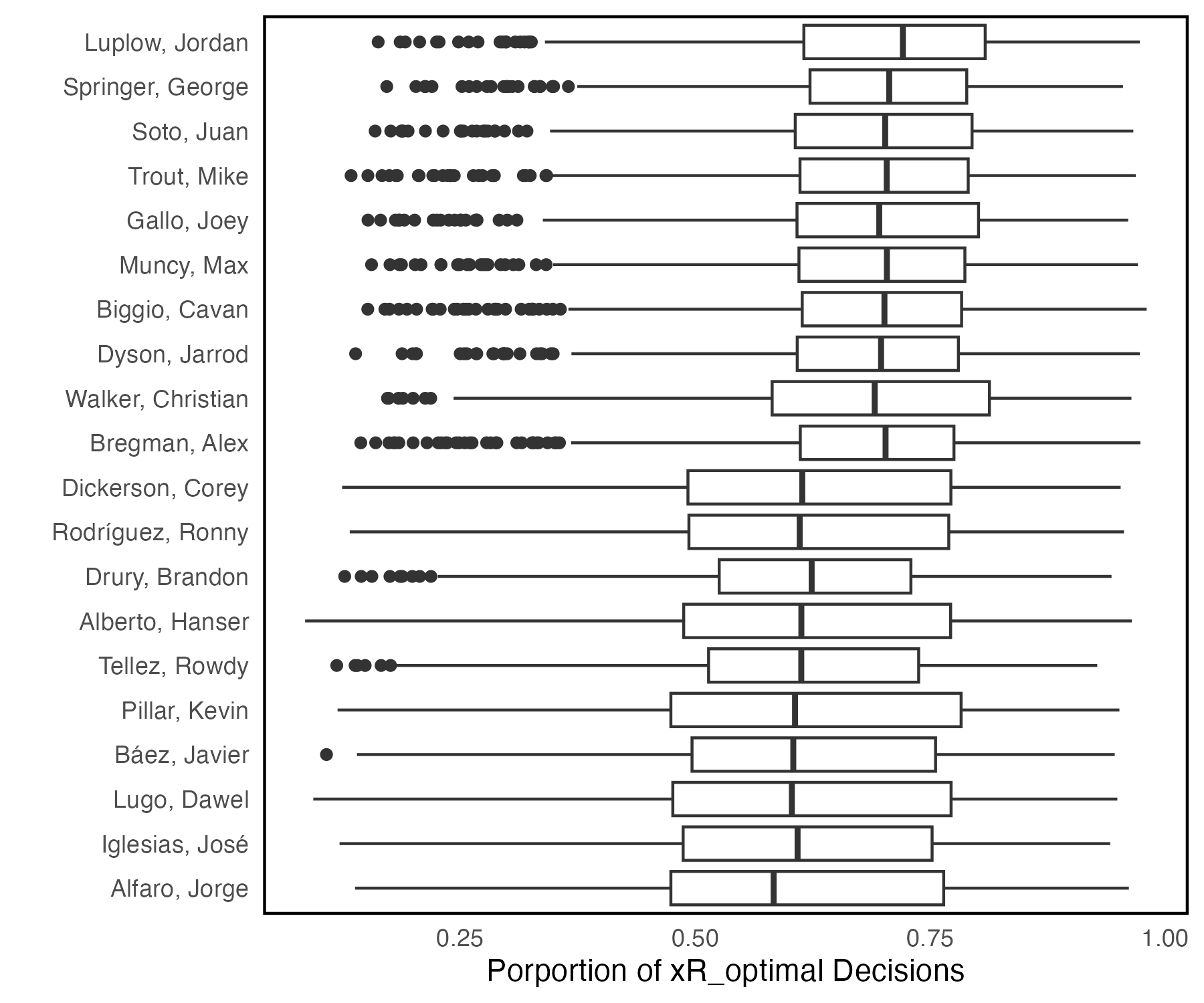}
    \caption{Boxplots of posterior samples of the proportion of pitches where the batter made the $\correct$ decision for the top 10 and bottom 10 batters who faced at least 1000 pitches in the 2019 MLB regular season.}
    \label{fig:prop_ci_boxplot}
\end{figure}

Figure~\ref{fig:prop_ci_boxplot} shows boxplots of the proportion of pitches where batter's made the $\correct$ decision for each posterior sample for the top 10 and bottom 10 batters who faced at least 1000 pitches in the 2019 MLB regular season.
There is considerable uncertainty in the proportion of pitches where batter's made the $\correct$ decision, and the overlap between batters demonstrates the difficulty in differentiating batters.

\begin{figure}[ht]
    \centering
    \includegraphics[width=.6\textwidth]{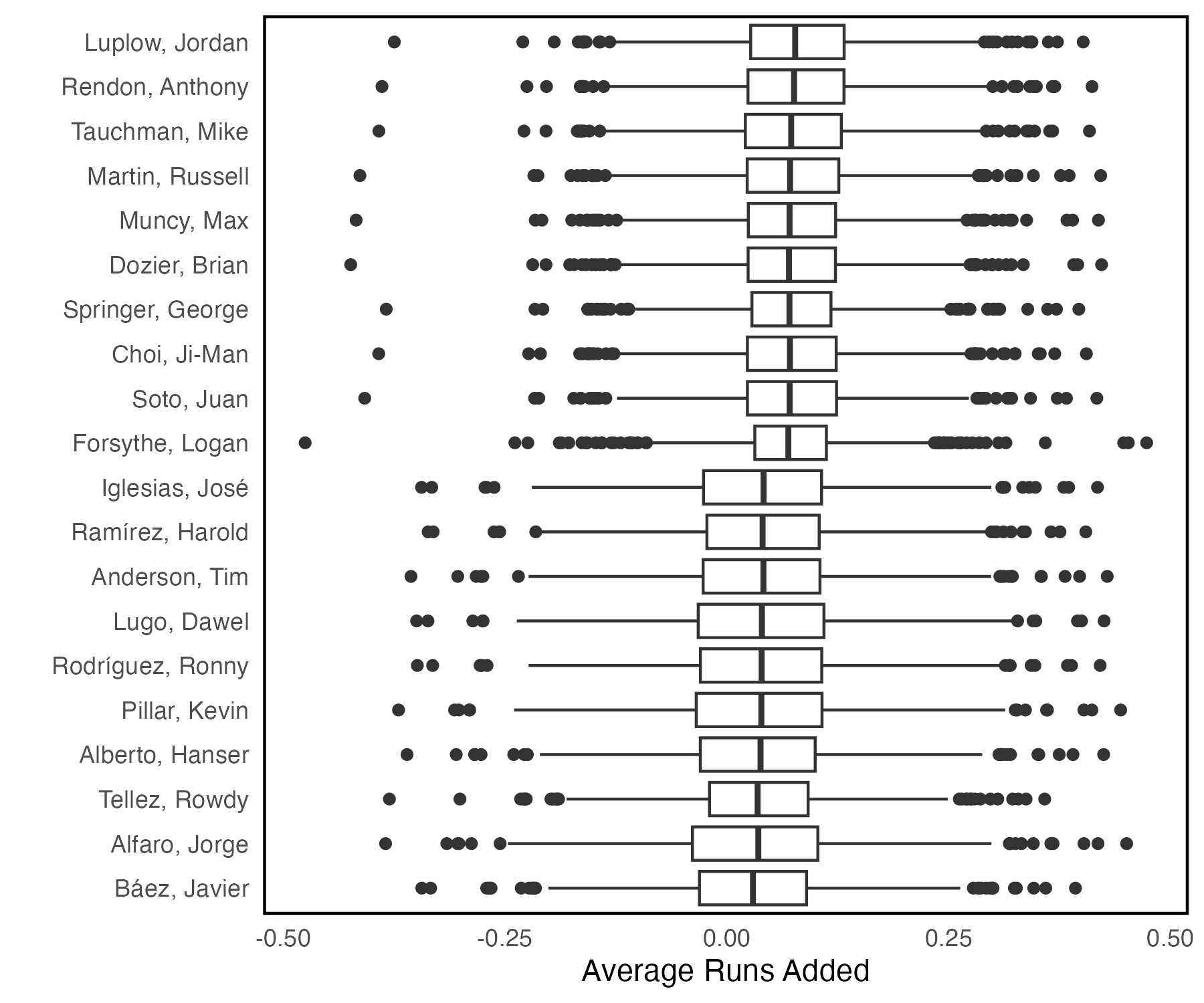}
    \caption{Boxplots of posterior samples of the runs added metric for the top 10 and bottom 10 batters who faced at least 1000 pitches in the 2019 MLB regular season.}
    \label{fig:added_ci_boxplot}
\end{figure}

Figure~\ref{fig:added_ci_boxplot} shows boxplots of the runs added metric for each posterior sample for the top 10 and bottom 10 batters who faced at least 1000 pitches in the 2019 MLB regular season.
The boxplots show considerable uncertainty in these estimates, and overlap in distributions between batters demonstrates the difficulty in differentiating batters based on this metric.

\begin{figure}[ht]
    \centering
    \includegraphics[width=.6\textwidth]{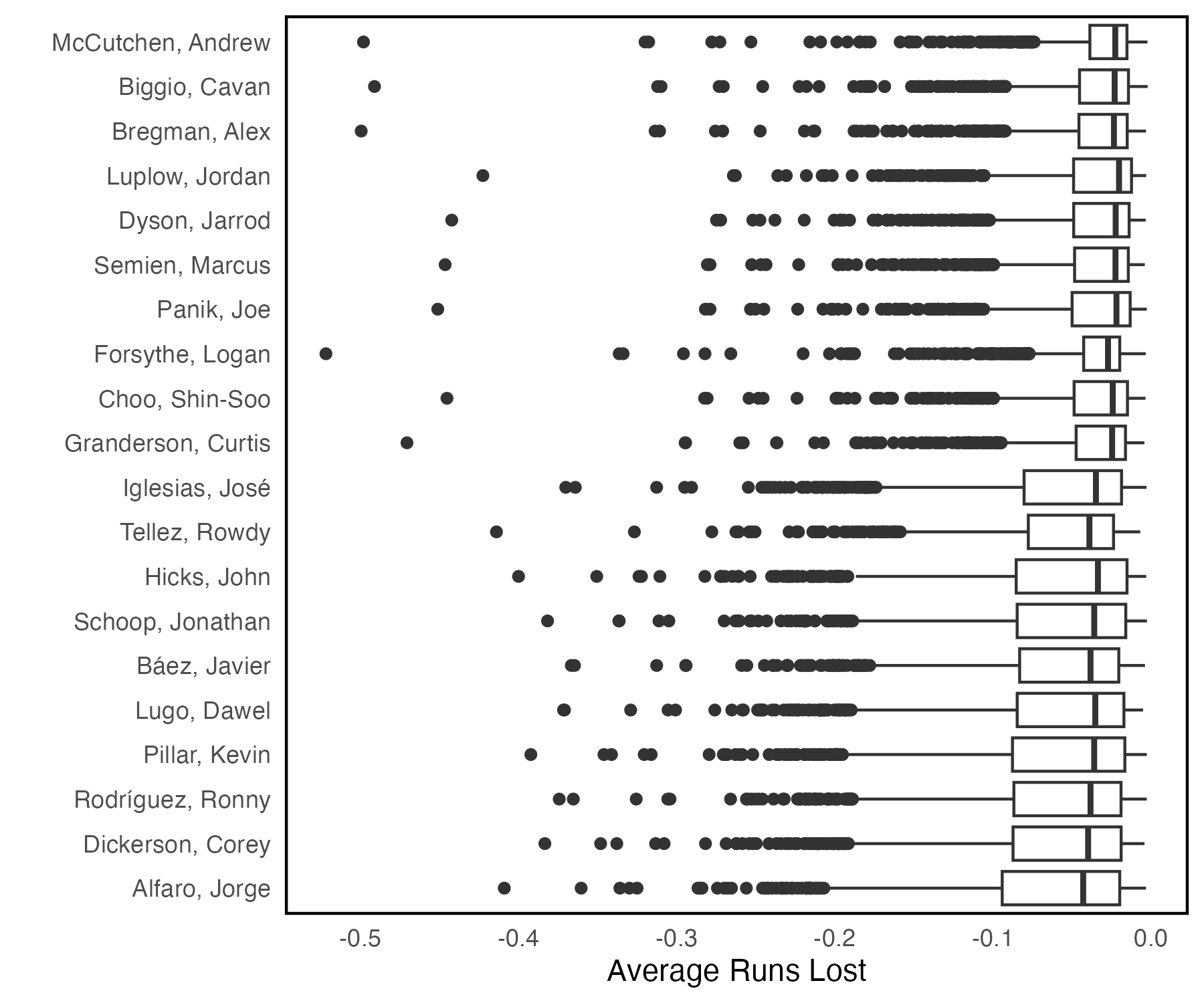}
    \caption{Boxplots of posterior samples of the runs lost metric for the top 10 and bottom 10 batters who faced at least 1000 pitches in the 2019 MLB regular season.}
    \label{fig:lost_ci_boxplot}
\end{figure}

Figure~\ref{fig:lost_ci_boxplot} shows boxplots of the runs lost metric for each posterior sample for the top 10 and bottom 10 batters who faced at least 1000 pitches in the 2019 MLB regular season.
There is considerable uncertainty in these estimates, and overlap in distributions between batters demonstrates the difficulty in differentiating batters based on this metric.

\end{document}